\documentclass[12pt]{article}
\pdfoutput=1
\usepackage{amsmath}
\usepackage{amssymb}
\usepackage{epsfig}
\usepackage{cite}
\usepackage{hyperref}
\usepackage{fontenc}
\usepackage{graphicx}
\usepackage{float}
\usepackage[lofdepth,lotdepth,caption=false]{subfig}
\usepackage{color}
\usepackage{booktabs}
\usepackage{tabularx}
\usepackage{multirow}
\usepackage{longtable}
\usepackage{lscape}
\usepackage{dcolumn}
\usepackage{rotating}%
\usepackage[normalem]{ulem}
\definecolor{darkgreen}{cmyk}{1,0,1,0.4}
\definecolor{brown}{cmyk}{0,0.8,1,0.2}
\definecolor{darkred}{cmyk}{0,1,1,0.2}
\allowdisplaybreaks[4] % for page break
\makeatletter
\renewcommand{\fnum@table}{\textbf{\tablename~\thetable}}
\renewcommand{\fnum@figure}{\textbf{\figurename~\thefigure}}
\makeatother

\newcounter{myenumi}

\renewcommand{\themyenumi}{\roman{myenumi}}
{\end{list}}

\newlength{\myem}
\newcounter{mysubequation}[equation]

\makeatletter
\renewcommand{\section}{\@startsection{section}{1}{0em}{-\baselineskip}%
{\baselineskip}{\normalfont\large\bfseries}}
\renewcommand{\subsection}%
{\@startsection{subsection}{2}{0em}{-0.7\baselineskip}%
{0.7\baselineskip}{\normalfont\bfseries}}
\makeatother

\newcommand{\bi}{\begin{itemize}}
\newcommand{\ei}{\end{itemize}}

\def\beq{\begin{equation}}
\def\eeq{\end{equation}}

\newcommand{\bea}{\begin{eqnarray}}
\newcommand{\eea}{\end{eqnarray}}

\newcommand{\pme}{P_{\mu e}}
\newcommand{\eet}{\varepsilon_{e\tau}}
\newcommand{\emt}{\varepsilon_{\mu\tau}}
\newcommand{\ett}{\varepsilon_{\tau\tau}}
\newcommand{\eee}{\varepsilon_{ee}}
\newcommand{\eem}{\varepsilon_{e\mu}}
\newcommand{\emm}{\varepsilon_{\mu\mu}}

\newcommand{\eetp}{\varphi_{e\tau}}
\newcommand{\emtp}{\varphi_{\mu\tau}}
\newcommand{\eemp}{\varphi_{e\mu}}
\def\epsilon{\varepsilon}

\newcommand{\chisq}{\ensuremath{\chi^2}}

\newcommand{\ttok}{{\sc T2K}}
\newcommand{\dune}{{\sc DUNE}}

\newcommand{\nova }{{\sc NOvA}}

\newcommand{\ttohk}{{\sc T2HK}}

\newcommand\sch{Schr$\ddot{\rm o}$dinger~}

\def\<{\langle}
\def\>{\rangle}

\def\dfrac#1#2{{\displaystyle\frac{#1}{#2}}}

\def\lsim{\mathrel{\rlap{\lower4pt\hbox{\hskip1pt$\sim$}}
    \raise1pt\hbox{$<$}}}         %less than or approx. symbol
\def\gsim{\mathrel{\rlap{\lower4pt\hbox{\hskip1pt$\sim$}}
    \raise1pt\hbox{$>$}}}         %greater than or approx. symbol

\begin{document}
%
%%%%%%%%%%%%%%%%%%%%%%%%%%%%%%%%%%%%%%%%%%%%%%%%%%%%%%%%%%%%%%%%%%%%%
%%%%                     Title-page                              %%%%
%%%%%%%%%%%%%%%%%%%%%%%%%%%%%%%%%%%%%%%%%%%%%%%%%%%%%%%%%%%%%%%%%%%%%

\begin{titlepage}

\renewcommand{\thefootnote}{\alph{footnote}}

\vspace*{-3.cm}
\begin{flushright}
CETUP2016-001\\
\end{flushright}
%\vspace*{0.5cm}
\renewcommand{\thefootnote}{\fnsymbol{footnote}}
\setcounter{footnote}{-1}
{
\begin{center}
{\large \bf
{{Non-standard interactions and the resolution of ordering of neutrino masses at DUNE and other long baseline experiments}
}
\\[0.2cm]
}
\end{center}
}

\renewcommand{\thefootnote}{\alph{footnote}}

\vspace*{.8cm}
\vspace*{.3cm}
{\begin{center} 
            {{\sf 
                Mehedi Masud$^\star$~\footnote[1]{\makebox[1.cm]{Email:}
                masud@hri.res.in}, and 
                Poonam Mehta$^{\ddagger}$~\footnote[2]{\makebox[1.cm]{Email:}
                pm@jnu.ac.in}
               }}
\end{center}}
\vspace*{0cm}
{\it 
\begin{center}
\footnotemark[1]%
$^\star$\, Harish-Chandra Research Institute, Chhatnag Road, Jhunsi, Allahabad 211 019, India
\footnotemark[2]%
$^\ddagger$ \, School of Physical Sciences, Jawaharlal Nehru University, 
      New Delhi 110067, India
\end{center}}

\vspace*{1.5cm}

\begin{center}
{\Large \today}
\end{center}

{\Large 
\bf
\begin{center} Abstract 
\end{center}  }
In the era of precision neutrino physics, we study the influence of matter NSI on the question of neutrino mass ordering and its resolution. At long baseline experiments, 
since matter effects play a crucial role  in addressing this very important question, it is timely to investigate how sub-leading effects due to NSI may affect and drastically 
alter inferences pertaining to this question. We demonstrate that the sensitivity to mass ordering gets significantly impacted due to NSI effects for various long baseline experiments  including  the upcoming long baseline experiment, Deep Underground Neutrino Experiment (DUNE). Finally we draw a comparison of DUNE, with the sensitivities offered by two of the current neutrino beam experiments NOvA and T2K.

 \vspace*{.5cm}

\end{titlepage}

\newpage

\renewcommand{\thefootnote}{\arabic{footnote}}
\setcounter{footnote}{0}

%~~~~~~~~~~~~~~~~~~~~~~~~~~~~~~~~~~~~~~~~~~~~~~~~~~~~~~~~~~~~~~~~~~~~~~~~~~~~~~~~~~~~~~~~~~~~~~~~~~~~~~~%

\section{Introduction}
\label{sec:intro}

A series of experiments using solar, atmospheric, accelerator and reactor neutrinos in the past several 
  decades have established beyond doubt that neutrinos oscillate among the three flavors while 
  conserving the lepton number. Oscillation experiments are sensitive to two mass-squared differences, 
  three mixing angles, and the value of the Dirac type CP violating phase, $\delta$.
 The latest global fit~\cite{Forero:2014bxa,Gonzalez-Garcia:2014bfa} to the world oscillation data 
 leads us to two intriguing aspects concerning mass and mixing in the neutrino sector.  
  The best-fit values~\cite{Forero:2014bxa,Gonzalez-Garcia:2014bfa} of the two mass-squared splittings and the angles are\footnote{The bracketted values correspond to IO.},
\begin{itemize}
          \item {\it Mass splittings  :}
          \begin{equation} \delta m^2_{21} =   7.5 \times 10^{-5}  eV^2~; \quad \delta m^2_{31} =
    2.457 (-2.449) \times 10^{-3}  eV^2~,
    \end{equation}     %
          \item {\it Mixing pattern :} 
          \begin{equation}
          \sin^2\theta_{12} = 0.304 ~;\quad 
          \sin^2\theta_{23} = 0.452 (0.579)  ~;\quad
          \sin^2\theta_{13} = 0.0218 (0.0219)~.
          \end{equation}
           \end{itemize}
       $ \delta m^2_{21} > 0$ is required from solar neutrino data while $ \delta m^2_{31}$ can be either positive or negative. So far there is no constraint on $\delta$, and it can lie in $[-\pi,\pi]$. Additionally if $\theta_
        {23}$ differs from maximal mixing as is hinted by recent global analyses of data, one would like to pin down the correct octant of this angle.         
        The emerging goals of neutrino oscillation physics are therefore to address 
  the question of neutrino mass ordering which refers to ascertaining whether $\delta m^2_{31} >0$  (normal ordering, NO) 
  or $\delta m^2_{31} < 0$ (inverted ordering, IO), to measure the CP violating phase and to find the correct
   octant of $\theta_{23}$ if $\theta_{23}$ turns out to be non-maximal.

It is crucial to settle the issue of neutrino mass ordering as it would allow us to get closer towards 
determining the underlying  structure of the neutrino mass matrix by being able to discriminate 
between theoretical models giving rise to neutrino masses~\cite{Albright:2006cw}. Knowledge of neutrino mass ordering
 would also  have an important bearing upon the neutrinoless double beta decay searches which would allow us to probe the nature of neutrinos~\cite{Pascoli:2005zb}. It is also intimately related to the measurement of CP violating phase, $\delta$. 
In a large class of theoretical models, the neutrino mass ordering also impacts the effectiveness of 
leptogenesis scenario which can explain the matter-antimatter asymmetry of the Universe~\cite{Fukugita:1986hr}.

Wolfenstein in his seminal paper~\cite{Wolfenstein:1977ue,Mikheev:1987qk} pointed out that  neutrinos could experience flavour-dependent refraction in matter which can modify the neutrino oscillation probabilities. In case of standard interactions (SI), the flavour dependent refraction arises due to the coherent forward scattering of $\nu_e$ with electrons present in matter via charged current (CC) processes while the NC contribution is flavour universal. However, in presence of NSI, one could have additional CC (see for example,~\cite{Ge:2016xya}) or  neutral current (NC) contributions. It turns out that only the NC processes play a role in the propagation of neutrinos and we will focus on these NC NSI in the present work.

The matter-induced modification of neutrino oscillation probabilities was found to be different for neutrino and antineutrino channels 
and a function of neutrino mass ordering. If we consider the $\nu_\mu \to \nu_e$ and $\bar\nu_\mu \to \bar
 \nu_e $ appearance channels, 
 one sees an enhancement in $P_{\mu e}$ and suppression in $\bar P_{\mu e}$ if ordering is normal 
 while if the ordering 
 is inverted, one gets the reverse. Since matter effects differ for neutrino and anti-neutrinos and 
 are sensitive to ordering, they aid in the determination of the neutrino mass ordering.  In presence of sub-dominant new physics effects such as non-standard interactions (NSI) during propagation, the resolution of neutrino mass ordering  gets severely affected even if we take conservative values of the NSI parameters (for reviews, see \cite{Ohlsson:2012kf,Miranda:2015dra}).

Recently, we have studied the influence of NSI during propagation (including both flavour non-universal and flavour changing interactions) on the CP violation sensitivity at a future super beam experiment, Deep Underground Neutrino  Experiment (\dune)~\cite{Marciano:2006uc,Bass:2013vcg,Acciarri:2015uup,Acciarri:2016ooe,Acciarri:2016crz} where we considered the so called {\bf{platinum channel}} ($\nu_\mu \to \nu_e$)~
\cite{Masud:2015xva} (see also~\cite{Coloma:2015kiu,deGouvea:2015ndi,Forero:2016cmb,Liao:2016hsa,Huitu:2016bmb,Bakhti:2016prn,Coloma:2016gei}). In Ref.~\cite{Masud:2016bvp}, we discussed the role of {\bf{disappearance 
channel ($\nu_\mu \to \nu_\mu$)}} in addition to the platinum channel and contrasted the CP sensitivities offered by some of the current and upcoming long baseline experiments such as Tokai to Kamioka (\ttok)~\cite{TheT2KCollaboration01042015}, NuMI off-axis $\nu_e$ appearance (\nova)~\cite{nova, Adamson:2016tbq}, \dune\, as well as  Tokai to Hyper-Kamiokande (\ttohk)~\cite{Abe:2015zbg}.

Apart from using the super beam experiments which are well-suited to exploit the platinum channel, the question of neutrino mass ordering can be addressed utilising sensitivity offered by other channels such as $\bar\nu_e \to \bar\nu_e$ in reactor experiments (e.g. Jiangmen Underground Neutrino Observatory  (JUNO)~\cite{Djurcic:2015vqa,An:2015jdp}), muon disappearance channel ($\nu_\mu \to \nu_\mu$) using atmospheric neutrinos in conjunction with an iron calorimeter detector (e.g. India-based Neutrino Observatory (INO)~\cite{Ahmed:2015jtv}),
  a combination of oscillation channels and exploring matter resonances in the multi megaton ice detector Oscillation Research with Cosmics in the Abyss (ORCA) or  Precision IceCube Next Generation Upgrade (PINGU)~\cite{Adrian-Martinez:2016fdl,Aartsen:2014oha,Ribordy:2013set,Akhmedov:2012ah} using atmospheric neutrinos and precision cosmology~\cite{Hannestad:2016fog}.

 In the present work, we include subdominant NSI effects (see~\cite{Antusch:2008tz,Farzan:2015doa,Farzan:2015hkd} 
 for models giving rise to NSI) and  evaluate the sensitivities and discovery potential offered by 
some of these long baseline accelerator experiments and assess their roles in 
addressing the question of neutrino mass ordering~\cite{Qian:2015waa}. We consider the following experiments : \ttok\, (295 km), \nova\, (800 km) and \dune\, (1300 km).

\ttok~\cite{TheT2KCollaboration01042015} and \nova~\cite{nova, Adamson:2016tbq}, are currently running  long baseline experiments while \dune~\cite{Marciano:2006uc,Bass:2013vcg,Acciarri:2015uup,Acciarri:2016ooe,Acciarri:2016crz} is one of the most promising upcoming long baseline experiments. The baseline of DUNE is chosen such that the experiment is expected to 
  deliver optimal sensitivity to CP violation and is well-suited to address 
  the question of neutrino mass ordering~\cite{Qian:2015waa}. The energy at which the
   neutrino flux peaks ($\sim 2.5$  GeV) and where the peak in $P_{\mu e}$ occurs match. 
In the region of peak flux, this asymmetry in neutrino and antineutrino channels is expected to
  be nearly 40\% which inturn implies that both the mass ordering and CP phase can be determined 
  unambiguously with high confidence within the same experiment.

While sensitivity studies have been carried out in presence of NSI in the context of \dune, we would like to stress that none of them deal with the precise impact of NSI on the standard sensitivity to mass ordering at long baseline experiments and it is timely to investigate extensively the impact of NSI on the question of mass ordering of neutrino states. We  consider NSI terms whose strengths lie in the presently allowed limits (along with the phases associated which are presently unconstrained) and study the impact of individual and collective NSI terms on the  the  CP violation sensitivity using the following channels : $\nu_\mu \to \nu_e$ and $\nu_\mu \to \nu_\mu$. In a recent work, the authors have discussed the mass ordering asymmetries 
in the $\nu_\mu \to \nu_\tau$ channel in presence of NSI and how it would impact the question 
of neutrino mass ordering at DUNE and the Long Baseline Neutrino Oscillations (LBNO) experiment~\cite{Rashed:2016rda}.

The paper is organised as follows. Sec.~\ref{sec:framework} gives the framework for the present work. Sec~\ref{framework_a} comprises of a brief introduction to NSI in propagation and how the NC NSI terms enter the oscillation framework.  We then review 
$\pme$ and $P_{\mu\mu}$ in Sec.~\ref{framework_b} and highlight the dependence on mass ordering as well as $\delta_{CP}$. 
In Sec.~\ref{framework_c}, we give our analysis procedure using the  dependence of probabilities on mass ordering in Sec.~\ref{framework_b}.
We then go on to describe our results in Sec.~\ref{results} where we show how mass 
ordering sensitivity at \dune\, gets affected due to individual and collective NSI terms 
(Sec.~\ref{results_b} and \ref{results_d}). We also show dependence on true 
values of standard oscillation parameters in Sec.~\ref{results_c} and compare the results obtained at \dune\, with other long baseline experiments in 
Sec.~\ref{results_d}. Using event rate plots for NO and IO, we depict 
degenerate and non-overlapping regions as a function of energy
 in Sec.~\ref{results_d}. 
The impact of NSI on the mass ordering fraction is shown as a 
function of exposure and baseline in Sec.~\ref{results_e} and \ref{results_f}.
We conclude with a discussion  in Sec.~\ref{sec:conclude}.

%~~~~~~~~~~~~~~~~~~~~~~~~~~~~~~~~~~~~~~~~~~~~~~~~~~~~~~~~~~~~~~~~~~~~~~~~~~~~~~~~~~~~~~~~~~~~~~~~~~~~~~~%
\section{Framework}
\label{sec:framework}
%-------------------------------------------
\subsection{Neutrino interactions and the Earth matter effects}
\label{framework_a}
%------------------------------------------- 
In presence of NSI, the propagation of neutrinos is governed by an effective \sch equation in the ultra-relativistic limit with the effective Hamiltonian  in flavour basis given by 
 \bea
 \label{hexpand} 
 {\mathcal
H}^{}_{\mathrm{f}} &=&   {\mathcal
H}^{}_{\mathrm{v} } +  {\mathcal
H}^{}_{\mathrm{SI} } +  {\mathcal
H}^{}_{\mathrm{NSI}} 
\nonumber 
\\
&
=&\lambda \left\{ {\mathcal U} \left(
\begin{array}{ccc}
0   &  &  \\  &  r_\lambda &   \\ 
 &  & 1 \\
\end{array} 
\right) {\mathcal U}^\dagger  + r_A   \left(
\begin{array}{ccc}
1  & 0 & 0 \\
0 &  0 & 0  \\ 
0 & 0 & 0 \\ 
\end{array} 
\right)  +
 {r_A}   \left(
\begin{array}{ccc}
\epsilon_{ee}  & \epsilon_{e \mu}  & 
\epsilon_{e \tau}  \\ {\epsilon_{e\mu} }^ \star & 
\epsilon_{\mu \mu} &   \epsilon_{\mu \tau} \\ 
{\epsilon_{e \tau}}^\star & {\epsilon_{\mu \tau}}^\star 
& \epsilon_{\tau \tau}\\
\end{array} 
\right) \right\}  \ ,
 \eea 
where 
\begin{equation}
\lambda \equiv \frac{\delta m^2_{31}}{2 E}  \quad \quad ; \quad \quad
r_{\lambda} \equiv \frac{\delta m^2_{21}}{\delta m^2_{31}} \quad \quad ; 
\quad \quad r_{A} \equiv \frac{A (x)}{\delta m^2_{31}} \ .
\label{dimless}
\end{equation}
 $A (x)= 2 \sqrt{2}  	E G_F n_e (x)$ is the standard CC potential due to
the coherent forward scattering of neutrinos in Earth matter and  $n_e$ is the electron
number density. 
${\mathcal
  U}$
is the three flavour neutrino mixing matrix and is responsible for diagonalizing the vacuum part of the Hamiltonian. 
In the Pontecorvo-Maki-Nakagawa-Sakata (PMNS) parametrization~\cite{Beringer:1900zz}, ${\cal U}$ is given by
\bea
{\mathcal U}^{} &=& \left(
\begin{array}{ccc}
c_{12} c_{13}   & s_{12} c_{13} & s_{13} e^{-i\delta} \\ -s_{12} c_{13} -c_{12} s_{13} s_{23} e^{i\delta} &
c_{12} c_{23}  -s_{12} s_{13} s_{23} e^{i\delta} & c_{13} s_{23}   \\ 
s_{12} s_{23} - c_{12} s_{13} c_{23} e^{i\delta}  & -c_{12} s_{23} - s_{12} s_{13} c_{23} e^{i\delta} & c_{13} c_{23} \\
\end{array} 
\right)   
 \ ,
\label{u}
 \eea 
where $s_{ij}=\sin {\theta_{ij}}$ and $c_{ij}=\cos \theta_{ij}$. 
Additionally, if neutrinos are Majorana particles, one can have 
two additional Majorana-type phases but those are irrelevant as far as neutrino oscillations are concerned.  In the standard paradigm, we note that there is only one parameter, the Dirac CP phase $\delta$ that is  responsible for genuine CP violating effects and SI with Earth matter can introduce additional 
  CP effects (referred to as fake CP effects) due to the fact that matter is CP asymmetric.

${\cal H}_{NSI}$ comprises of  off-diagonal ($\alpha \neq \beta$)  and diagonal  ($\alpha=\beta$) parameters.
The off-diagonal NSI parameters, ${\varepsilon}_{\alpha \beta} \, (\equiv |\varepsilon _{\alpha \beta}|\, e^{i \varphi_{\alpha\beta}})$ are complex while the diagonal ones are real due to the
 hermiticity of the Hamiltonian. 
In addition to $\delta$ appearing in ${\cal H}_{\rm f}$, we now have  three other phases as $\eemp,\eetp,\emtp$.

%It is interesting to note that matter (or propagation) NSI obey unitarity (while source and detector NSI do not~\cite{Khan:2016uon,Ge:2016xya}) so effectively we still have an overall unitary matrix that diagonalises the effective Hamiltonian in presence of matter NSI.

Let us briefly mention the constraints on the NC NSI parameters (for more details, see Ref.~\cite{Chatterjee:2014gxa,Choubey:2015xha,Ohlsson:2012kf}). We use the following constraints on the NC NSI parameters 
 \begin{eqnarray}
 |\varepsilon_{\alpha\beta}|
 \;<\;
  \left( \begin{array}{ccc}
4.2  &
0.3 & 
0.5 \\
0.3 & 0.068 & 0.04 \\
0.5  &
0.04 &
0.15 \\
  \end{array} \right) \ .\label{tinynsi}
\end{eqnarray} 
The phases are unconstrained presently and can lie the allowed range, $\varphi_{\alpha\beta} \in (-\pi,\pi)$ (see Table.~\ref{tab:parameters}).

%-------------------------------------------
\subsection {Probability level discussion : dependence on the neutrino mass 
ordering and the CP phase at long baseline experiments}
\label{framework_b}
%-------------------------------------------

Let us first describe the explicit dependence of probabilities on two of the unknowns - neutrino mass ordering and CP phase \footnote{For non-maximal $\theta_{23}$, the octant of $\theta_{23}$ also needs to be taken into account. Here we take $\theta_{23} = \pi/4$ for simplicity.}. As we will see, the dependence of probabilities on these two parameters is strongly interlinked. 
We discuss the appearance ($\nu_\mu \to \nu_e$) and disappearance ($\nu_\mu\to\nu_\mu$) channels that are relevant for accelerator-based  neutrino oscillation experiments. 
The  expressions given below are strictly valid when $r_\lambda \lambda L/2 \ll 1$  i.e. $L$ and $E$ are far away from the region where lower frequency oscillations dominate which is generally satisfied for long baseline experiments. For the case of DUNE, we have 
\begin{equation}
 r_\lambda \lambda L/2 = 0.05 \,\left[1.267 \times \dfrac{\delta m^2_{21}}{7.6 \times 10^{-5} ~eV^2} \dfrac{L}{1300~km} \dfrac{2.5~GeV}{E} \right] < 1
 \label{lowfreq}
\end{equation}
 For the case  SI with matter (as well as in vacuum 
in the limit, $r_A \to 0$) we can express the probabilities for NO and IO in a 
compact form as given below (see \cite{Akhmedov:2004ny})  
%
%%%%%%%%%%%%%%%%%%%%%%%%%%%%%%
\begin{enumerate}
\item Normal mass ordering  : 
\begin{subequations}
\bea
P_{\mu e}^{NO} &=& x^2 + y^2 + 2 x y \cos (\delta + \lambda L/2)~,
\label{pno1}
\\
\bar P_{\mu  e}^{NO} &=& \bar x ^2 +  y^2 + 2 \bar x y \cos (\delta-\lambda L/2)~, 
\label{pno2}
\\
P_{\mu \mu}^{NO} &=&  a+b+c+e-y^2-d^2 - 2 y d \cos \delta~, 
\label{pno3}
\\
\bar P_{\mu  \mu}^{NO} &=& a+b+\bar c + \bar e -y^2 -\bar d^2 - 2 y \bar d \cos\delta~,
\label{pno4}
\eea
\end{subequations}
\item Inverted mass ordering   : 
\begin{subequations}
\bea
P_{\mu e}^{IO} &=& \bar x^2 + y^2 - 2 \bar x y \cos (\delta - \lambda L/2)~,
\label{pio1}
 \\
\bar P_{\mu  e}^{IO} &=& x ^2 +  y^2 - 2  x y \cos (\delta + \lambda L/2)~,
\label{pio2} \\
P_{\mu \mu}^{IO} &=& a - b + \bar c +\bar e -y^2 -\bar d^2 + 2 y \bar d \cos \delta
 \label{pio3} \\
\bar P_{\mu  \mu}^{IO} &=& a - b + c + e - y^2 -d^2 + 2 y d \cos\delta~,
\label{pio4}
\eea
\end{subequations}
where 
\begin{eqnarray}
x &=& s_{2\times {13}} s_{23} \frac{\sin \{{(1-r_A) \lambda L}/
2\}}{(1-r_A)}~,
\nonumber\\
y&=&r_\lambda s_
{2 \times {12}} c_{23} \frac{\sin ({r_A \lambda L}/2)} {r_A}
\nonumber\\
a &=& 1-s^2_{2\times 23} \sin^2 (\lambda L/2) -r_\lambda^2 c_{12}^4 s^2_{2 \times 23} (\lambda L / 2)^2 \cos \lambda L~,
\nonumber\\
 b&=& r_\lambda c_{12}^2 s^2_{2\times 23} (\lambda L/2) \sin \lambda L~,
 \nonumber\\
 c&=&\frac{1}{2r_A} r_\lambda^2 s^2_{2 \times 12} s^2 _{2\times 23} \big[ \sin (\lambda L/2) \frac{\sin (r_A \lambda L/2)}{r_A} \cos \{(1-r_A) \lambda L / 2\} - (\lambda L /4) \sin \lambda L\big]~,
 \nonumber\\
 d &=& 2 s_{13} s_{23} \frac{\sin (1-r_A)\lambda L/2}{1-r_A}~,
 \nonumber\\
 e &=&\frac{2}{1-r_A} s_{13}^2 s_{2 \times 23} ^2 \big[ \sin (\lambda L/2) \cos (r_A \lambda L/2) \frac{\sin \{(1-r_A)\lambda L/2\}}{1-r_A} - (r_{A} \lambda L / 4) \sin \lambda L\big]~,\nonumber \\ \end{eqnarray}
where $s_{ij} = \sin \theta_{ij}$, $c_{ij} = \cos \theta_{ij}$, $s_{2 \times ij} = \sin 2\theta_{ij}$
 with $i,j=1,2,3$. 
 %
%-------------------------------------------
\begin{figure}[ht]
\centering
\includegraphics[width=\textwidth]{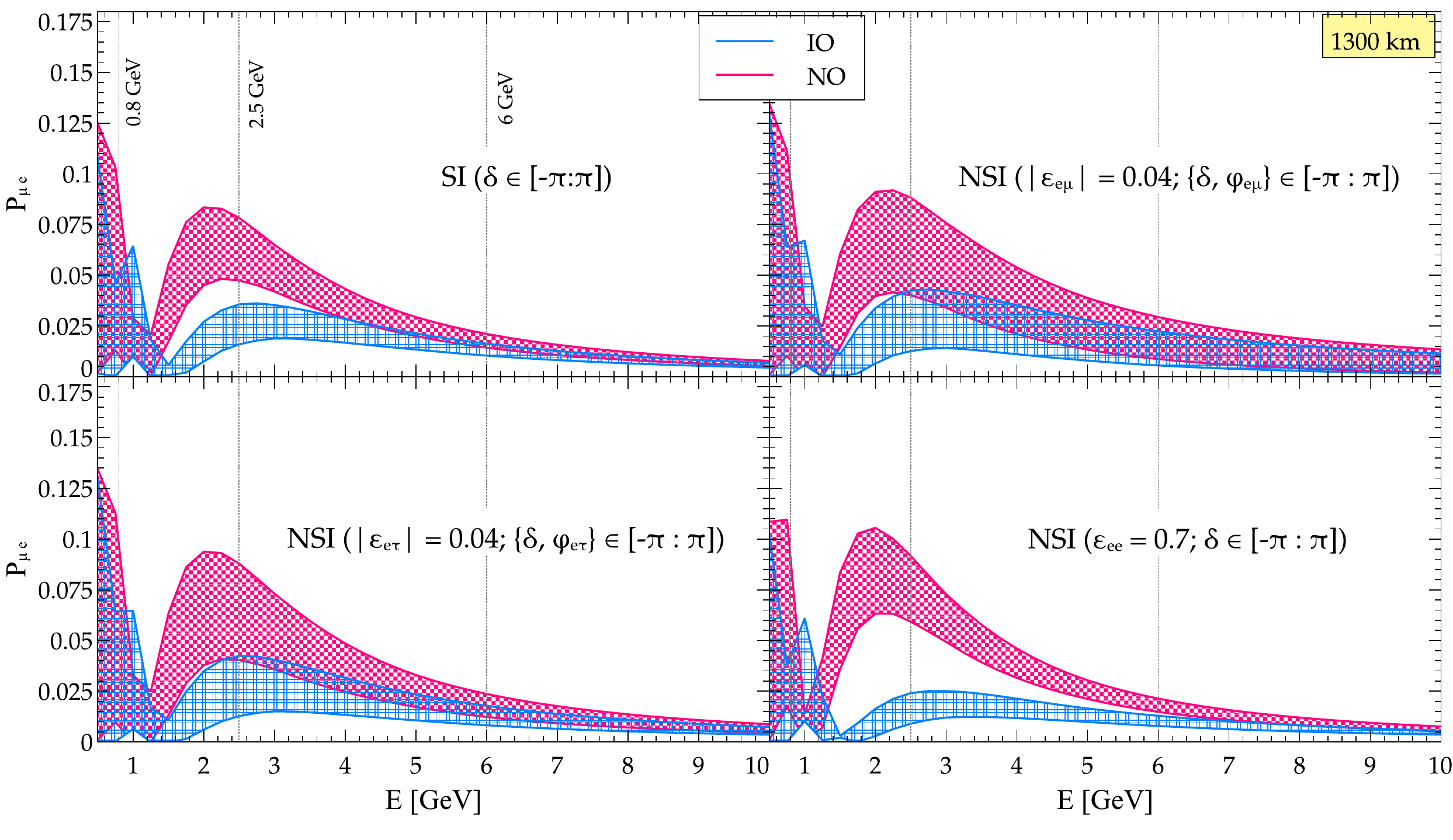}
\caption{\footnotesize{$P_{\mu e}$  as a function of $E$ for \dune\,  in presence of
 SI and NSI (considering one NSI parameter non-zero at a time) for the two choices of orderings.  
 }}
\label{fig0}
\end{figure}
%-------------------------------------------
%
 In addition to the vacuum oscillation frequency $\lambda L/2$ given by  
\begin{equation}
\lambda L/2 \approx 1.57 \, \left[1.267 \times \dfrac{\delta m^2_{31}}{2.5 \times 10^{-3} ~eV^2} \dfrac{L}{1300~km} \dfrac{2.5~GeV}{E} \right]~, 
\label{vpeak}
\end{equation}
which is $E$-dependent,  matter (SI and NSI) introduces phase shifts such as $r_A  \lambda L/2$ 
\begin{equation}
r_A \lambda L/2 \approx 0.4 \, \left[ 1.267 \times {0.756 \times 10^{-4}} \, \dfrac{\rho}{3.0 ~g/cc}\, \dfrac{L}{1300~km}\right]~,
\label{matterterm}
\end{equation} 
which are $E$-independent. 
 Given the probability expressions for the neutrino channel for a given ordering of neutrino mass states, we can get the corresponding expressions for the antineutrino channel by replacing $\delta \to -\delta$ and 
 $r_A \to - r_A$. For  IO, we need to substitute  $r_A \to - r_A$, $\lambda \to -
 \lambda$ and $r_\lambda \to - r_\lambda$ in the expression for NO. 
 \end{enumerate}

In what follows, we shall  see that Eqs.~\ref{pno1}-\ref{pno4} and Eqs.~\ref{pio1}-\ref{pio4} serve as useful guide to understand the impact of neutrino mass ordering.

As far as $P_{\mu e}$ is concerned (see Eqs.~\ref{pno1}-\ref{pno2} and Eqs.~\ref{pio1}-\ref{pio2}), $x$ is the dominant contributor  which depends upon the choice of the neutrino mass
 ordering. Close to the 1-3 resonance condition~\footnote{The resonance condition is sensitive to the mass ordering (NO or IO), for antineutrinos resonance occurs for IO. The dependence on $\delta m^2_{31}$   is strong while the dependence on $\theta_{13}$ is very mild.},
 \begin{equation}
\rho E_R [GeV ~ g/cc]  = \frac{\delta m^2_{31} [eV^2]}{0.76 \times 10^{-4} [eV^2]} 
\times \cos 2\theta_{13} 
\simeq 30 ~GeV ~ g/cc~,
 \end{equation}
  this term can be especially  enhanced due to matter effects since $r_A \to 1$. 
    $y$ on the other hand, is suppressed by $r_\lambda$.
    %
%-------------------------------------------
\begin{figure}[htb]
\centering
\includegraphics[width=\textwidth]{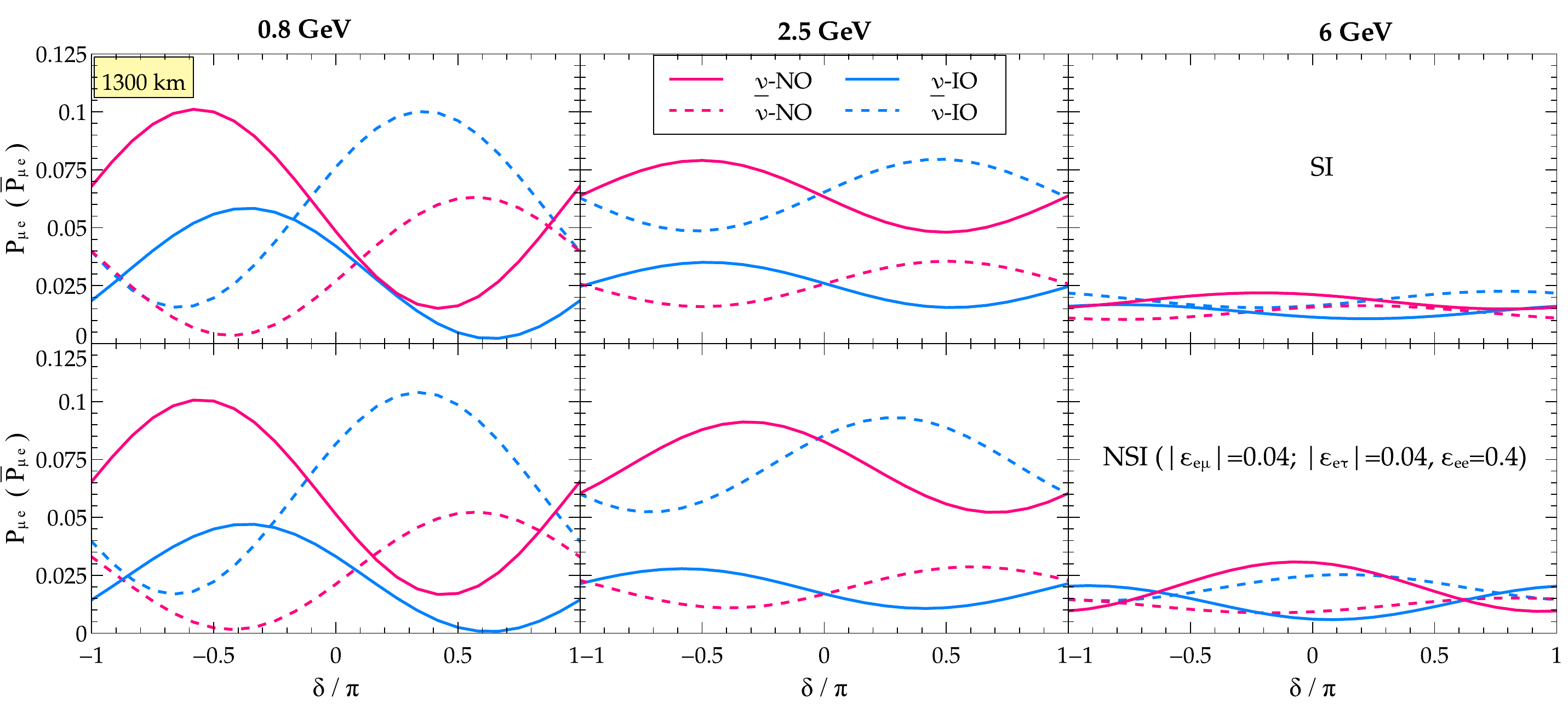}
\caption{\footnotesize{$P_{\mu e} (\bar P_{\mu e})$ as a function of the CP phase, $\delta$ for three values of energy ($E=0.8,2.5,6$ GeV) and fixed baseline ($L=1300$ km) in presence of SI (top row) and NSI (bottom row).  %
 }}
\label{fig1}
\end{figure}
%-------------------------------------------
 %
     The interference terms in Eqs.~\ref{pno1}-\ref{pno2} and Eqs.~\ref{pio1}-\ref{pio2} involving $x$ and $y$ depend
     upon two of the unknowns - the CP phase ($\delta$) as well as the choice of  
     mass ordering (via the $\lambda$ term). The interplay of these two dependencies can lead to degenerate solutions.

In Fig.~\ref{fig0}, for a fixed $L$($=1300$ km), we depict the $\nu_e$ appearance probability as a function of energy. The two bands correspond to variation in the values of phases\footnote{$\delta$ for SI and also relevant $\varphi_{\alpha\beta}$ for NSI} 
for NO and IO. 
 For the  SI case, we note that  the two bands corresponding to NO and IO are non-overlapping 
  in the energy range around the position of the first vacuum peak (around $E \sim 2.5$ GeV, see Eq.~\ref{vpeak}) 
  of $P_{\mu e}$. The top boundary of the band corresponds to $\delta \sim -\pi/2$ for NO and IO while the bottom boundary of the band corresponds to $\delta \sim \pi/2$ (see Eq.~\ref{pno1} and \ref{pio1}). 
   For the NSI case, if we consider $\varepsilon_{ee} > 0$, the degree of separation between 
   the two bands is more in comparison to SI case and also this extends to larger range of energies around the  peak of $P_{\mu e}$. 
If we consider $\varepsilon_{e\mu}$ and $\varepsilon_{e\tau}$, we note that 
  the two bands are overlapping at all energies 
 with the degree of overlap decreasing around the vacuum peak 
 (for similar discussion for the SI case in the context of NoVA and T2K see~\cite{Prakash:2012az}).
This naively tells us that for \dune\, the region around $2.5$ GeV is the most relevant 
  region for an unambiguous determination of
   mass ordering as the overlap between the NO and IO bands is minimal around this energy. 

  Let us now visualize the dependence of probability 
  as a function of $\delta$ for fixed $E$ and $L$ as shown in Fig.~\ref{fig1}. 
  Our choice of fixed  energies  correspond to the first vacuum peak ($2.5$ GeV), second vacuum peak 
  ($0.8$ GeV) and a higher value of energy beyond the first peak ($6$ GeV).
  An understanding of the
   explicit $\delta$ dependence will help us interpret the plots for sensitivity to mass ordering
     which are usually plotted as a function of the unknown standard Dirac CP phase  $\delta$.   
  The solid (dashed)  magenta  and blue curves are for neutrinos (antineutrinos) for NO and IO 
  respectively.
  The SI case is shown in the top row for three different fixed values of energies while the NSI case is depicted in the bottom row for those specific choices of energies. 
 At $E = 0.8$ GeV from Fig.~\ref{fig0}, we see that the two bands corresponding to NO and IO are
  overlapping. 
  %
 %-------------------------------------------
\begin{figure}[htb]
\centering
\includegraphics[width=\textwidth]{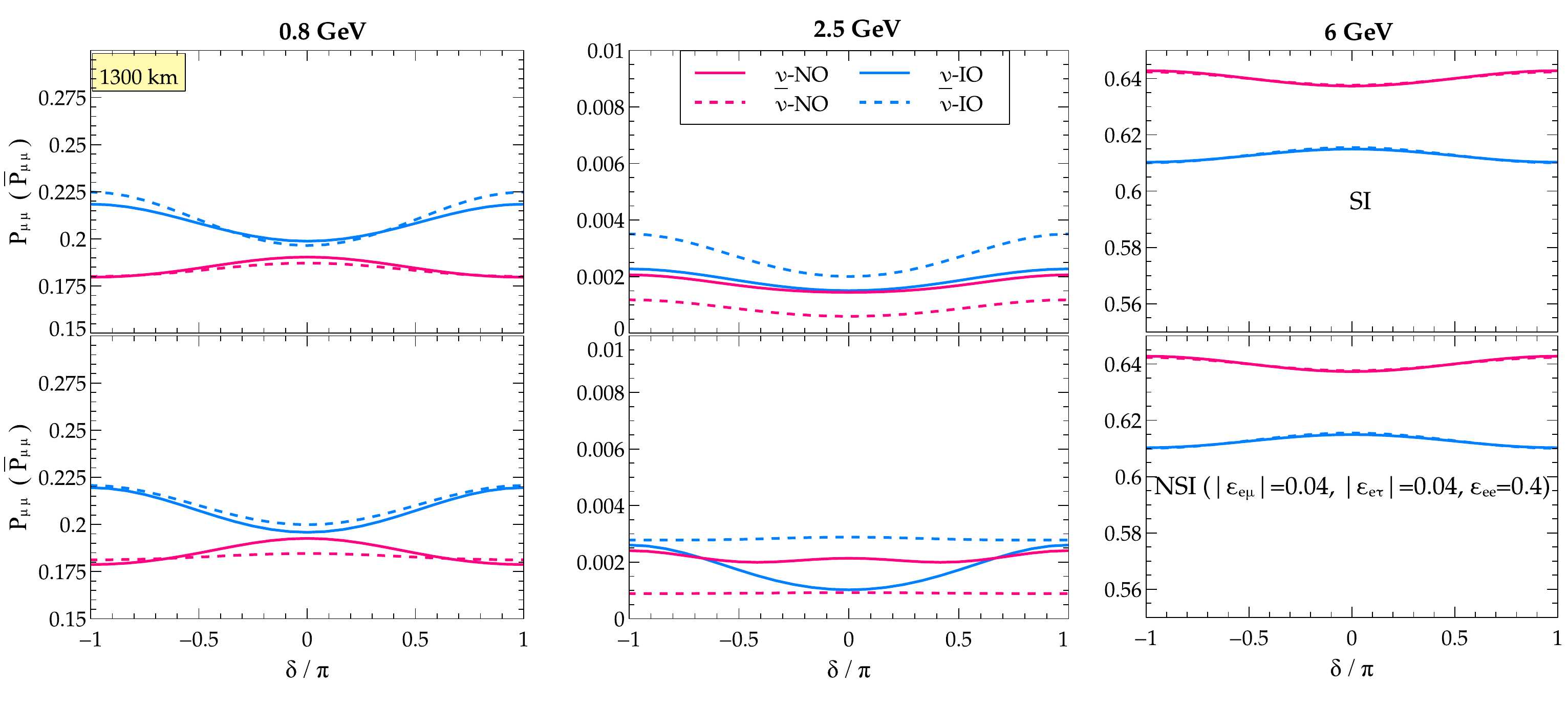}
\caption{\footnotesize{
$P_{\mu \mu} (\bar P_{\mu \mu})$ as a function of the CP phase, $\delta$ for three values of energy ($E=0.8,2.5,6$ GeV) and fixed baseline ($L=1300$ km) in presence of SI (top row) and NSI (bottom row).
 }}
\label{fig2}
\end{figure}
%-------------------------------------------
%
   This can also be seen in Fig.~\ref{fig1} (top row, left plot) as the red and 
   blue curves touch each other   on the 
  $\delta > 0$ side. 
  At $E=2.5$ GeV from Fig.~\ref{fig0}, the two bands are non-overlapping and therefore we notice that 
  the red and the blue curves  are not crossing over in Fig.~\ref{fig1} (top row, middle plot) 
  for all values of $\delta$. For higher energies $E = 6$ GeV, we can see that there is an
   overlap in Fig.~\ref{fig0} and therefore the red and the blue curves are crossing over at two 
   points on both sides of $\delta=0$ (top row, right panel of Fig.\ \ref{fig1}).
   
   Also, note that in Eqs.~\ref{pno1}-\ref{pio4}, the two frequencies involved are : 
   $\lambda L/2 $ which is energy-dependent and $r_A \lambda L/2$ which is energy-independent (see 
   Eq.~\ref{vpeak} and \ref{matterterm}). 
   By comparing the relative oscillation
 frequencies, we note that the energy-independent contribution should be the same for all energies
  and the $E$- dependent contribution is expected to rise with the decrease in energy for a fixed value of $L$. This is evident from Fig.~\ref{fig1} as we go from right to left.
 
As far as $P_{\mu\mu}$ is concerned, the plots versus $\delta$ are shown in Fig.~\ref{fig2}
and the features can be understood from Eqs.~\ref{pno3} and \ref{pio3}.
 The  peak (dip) conditions for $P_{\mu \mu}$ are shifted in $\delta$ w.r.t. 
$P_{\mu e}$. The amount of this shift depends upon the value of the energy and 
the baseline involved. Around $2.5$ GeV, we note that the position of peak 
of $P_{\mu\mu}$ is shifted by $\pi/2$ w.r.t. $P_{\mu e}$. At a given energy, we can see the value of 
 $P_{\mu\mu}$ is highest around $6$ GeV and not at $2.5$ GeV where the $P_{\mu e}$ 
  takes its maximum value and therefore one would expect that the contribution will be dominated by 
  $P_{\mu e}$ as the flux is the largest there. 
 Overall, from Fig.~\ref{fig2}, we note that  $P_{\mu \mu}$ is not very sensitive to NSI effects (the curves in the top and bottom panel are similar). We can see that the curves for  NO and IO in the bottom panel (NSI) of Fig.~\ref{fig2} appear different only (small relative difference) in the case of $2.5$ GeV but one should notice that 
 the overall value of probability scale is smaller by at least two orders of magnitude in this case. 
 The SI and NSI curves for NO and IO are very similar both at $0.8$ and $6$ GeV.
 Nevertheless the small contribution from $P_{\mu\mu}$ can be understood from the plots given in Fig.~\ref{fig2}.

%-------------------------------------------
\begin{table}[htb]
\centering
\begin{tabular}{ |l l l|}
\hline
&&\\
Parameter & True value & Marginalisation range  \\
&&\\
\hline
&&\\
{\sl{SI}} &&\\
&&\\
$\theta_{12}$ [deg] & 33.5  &  - \\
$\theta_{13}$ [deg] & 8.5 & -\\
$\theta_{23}$ [deg] & 45  & - \\
$\delta m^2_{21}$ [$eV^2$]  & $7.5 \times 10^{-5}$ & - \\
$\delta m^2_{31}$ (NH) [$eV^2$] & $+2.45 \times 10^{-3}$ & $+(2.25 - 2.65) \times 10^{-3}$ \\
$\delta m^2_{31}$ (IH) [$eV^2$] & $-2.46 \times 10^{-3}$ &  $-(2.25 - 2.65) \times 10^{-3}$  \\
$\delta$ & $ [-\pi : \pi]$ & $[-\pi : \pi]$\\
&&\\
\hline
&&\\
{\sl{NSI}} &&\\
&&\\
$\eee$ & $\pm 0.1,\pm 0.4,\pm 0.7$ & $[-0.7 : 0.7]$ \\
$\emm$ & $\pm 0.05$ & $[-0.06 : 0.06]$ \\
$\ett$ & $\pm 0.04,\pm 0.08, \pm 0.12$ & $ [-0.15 : 0.15]$ \\
$|\eem|$ & $0.01,0.04,0.07$ & $[0:0.10]$ \\
$|\eet|$ & $0.01,0.04,0.07$ & $[0:0.10]$ \\
$|\emt|$ & $0.01,0.04$ & $[0:0.04]$\\
$\phi_{e\mu}$ & $ [-\pi : \pi]$ & $ [-\pi : \pi]$\\
$\phi_{e\tau}$ & $ [-\pi : \pi]$ & $[-\pi : \pi]$\\
$\phi_{\mu\tau}$ & $ [-\pi : \pi]$ & $[-\pi : \pi ]$\\
\hline
\end{tabular}
\caption{\label{tab:parameters}
 SI and NSI parameters used in our study. For latest global fit to neutrino data see~\cite{Gonzalez-Garcia:2014bfa}.}
\end{table}
%-------------------------------------------
All the plots presented in this paper are obtained by using General Long baseline Experiment Simulator (GLoBES) and related software~\cite{Huber:2004ka,Kopp:2006wp,Huber:2007ji,Kopp:2007ne} which numerically 
solves  the full three flavour neutrino propagation equations using the PREM~\cite{Dziewonski:1981xy} density profile of the Earth\footnote{We use the matter density as given by PREM model.  
In principle, we can allow for uncertainity in the Earth matter density in our calculations but  it would not impact our results drastically~\cite{Gandhi:2004bj}.}, and the latest values of the neutrino parameters as obtained from global
fits~\cite{GonzalezGarcia:2012sz,Capozzi:2013csa,Forero:2014bxa}. 
 Unless stated otherwise, we assume NO as the true hierarchy in all the plots.

%-------------------------------------------
  \subsection{Analysis procedure and interpretations based on probability expressions 
  }
  \label{framework_c}
  %-------------------------------------------
    
The question of  ordering of neutrino mass eigenstates is a binary one : the true ordering can be NO or IO.
In order to obtain the sensitivity to neutrino mass ordering
 we need to ask the following question - 
what is the sensitivity with which a particular experiment can 
distinguish between NO and IO. We explore this question as a function of the 
Dirac CP phase $\delta$ as it is the  only unknown parameter in the PMNS mixing matrix. 

In order to understand the features of the sensitivity plots considering the true ordering as NO, we give a  statistical definition of $\chi^2$ as follows~\footnote{$N_\sigma = \sqrt{\Delta \chi^2}$. 
$\Delta \chi^2 = \chi^2$ as we have not included any fluctuations 
in simulated data. 
This is the Pearson's definition of $\chi^2$~\cite{Qian:2012zn}.  For large sample size, the other 
 definition using log-likelihood also yields similar results.} 
\begin{equation}
\label{chisq}
\chi^2_{NO} (\delta_{tr},|\varepsilon_{tr}|,\varphi_{tr})\equiv  \min_{\delta_{te},
|\varepsilon_{te}|,\varphi_{te}}  \sum_{i=1}^{x}  \sum_{j=1}^{2} \sum_{k=1}^2
 \frac{\left[N_{NO}^{i,j,k}(\delta_{tr},|\varepsilon_{tr}|,\varphi_{tr}) - 
 N_{IO}^{i,j,k} (\delta_{te} , |\varepsilon_{te}| , \varphi_{te} )\right]^2 }
 {N_{NO}^{i,j,k} (\delta_{tr},|\varepsilon_{tr}|,\varphi_{tr})}~,
\end{equation}
where $N_{NO}^{i,j,k}$ and $N_{IO}^{i,j,k}$ are the number of NO and IO events in the $\{i,j,k\}$-th bin respectively. The NSI parameters are expressed in terms of moduli, $|\varepsilon| \equiv \{ |\varepsilon_{\alpha\beta}| ; \alpha,\beta = e,\mu,\tau\}$ and phases, $\varphi \equiv \{\varphi_{\alpha\beta} ; \alpha,\beta = e,\mu,\tau \}$. 
The indices $i,j$ correspond to energy bins  ($i=1 \to x$, the number of bins depends upon the particular experiment - for \dune, there are $x=39$ bins of width 250 MeV in $0.5-10$ GeV, for \ttok\, and \ttohk, there are $x=20$ bins of width 40 MeV  in $0.4-1.2$~GeV, for \nova, there are $x=28$ bins of width 125 MeV in $0.5-4$ GeV) and the type of neutrinos i.e. neutrino or antineutrino ($j = 1 \to 2$).  
$k$ stands for the channels considered i.e. appearance and disappearance ($k = 1 \to 2$). 
The $\chi^2$ is computed~\footnote{Note that this definition suffices to illustrate the behaviour of $\chi^2$ in terms of various terms appearing in the probability expression. For all the actual $\chi^{2}$ results we have used GLoBES which uses Poissonian definition of $\chi^2$.} as given in Eq.~\ref{chisq} for a given set of true values by minimizing over the test parameters and this procedure is repeated for all possible true values  listed in Table~\ref{tab:parameters}. 
We do not marginalise over the standard oscillation parameters except $\delta$ whose true value is unknown. 
As we are investigating the role of NSI in the present study, we marginalize over the allowed ranges of moduli and phases of the relevant NSI parameters. Our choice of range of NSI parameters is consistent with 
the  existing constraints (Eq.~\ref{tinynsi}). 
  While the variation corresponding to the true value of $\delta$  is depicted 
  along the $x-$axis, the variation  of the true values of $\varphi_{e\mu},\varphi_{e\tau}$ (within the allowed ranges) lead to the vertical width of the sensitivity bands which show the maximum variation in the $\chi^2$ for each value of $\delta$ (true)\footnote{The true values of $\varphi_{e\mu}$ and $\varphi_{e\tau}$ that lead to maximum and minimum $\chi^2$ are in general not the same for each $\delta$ (true). }.

For  the sake of clarity, we have retained only statistical effects and ignored systematic uncertainties and priors  in the above expression in eq.\ \ref{chisq} (see~\cite{Barger:2014dfa} for the full expression of $\chi^2$ including systematics and priors). It turns out  that the above expression suffices to explain the gross behaviour of $\chi^2$ curves in connection to  the issue of mass ordering. In practice, we have indeed included the  effects due to systematics for different experiments and also marginalised over systematic uncertainties while calculating the $\chisq$ using GLoBES. 
We have assumed that the standard oscillation parameters are known with infinite precision i.e. we have not included any priors.  

The theoretically expected differential event rate is given by~\cite{Bass:2013vcg},  
\bea
\label{eq:eventrate}
\frac{dN^{app}_{\nu_e} (E,L)}{dE}  &=& N_{target}  \times 
 \Phi_{\nu_\mu} (E,L)  \times P_{\mu e} (E,L) \times \sigma_{\nu_e} (E) ~,
 \eea where
  $N_{target}$ is the number of target nucleons per kiloton of detector fiducial volume, $N_{target}=6.022 \times 10^{32}~N/kt$. 
  $P_{\mu e}  (E,L)$ is the appearance probability for $\nu_\mu \to \nu_e$ in matter, $\Phi_{\nu_\mu} (E,L)$ is the flux of $\nu_\mu$, 
 $\sigma _{\nu_e} (E)$ is the charged current (CC) cross section of $\nu_e$ given by 
 \begin{equation}
 \sigma_{\nu_e} = 0.67 \times 10^{-42} (m^2/GeV/N) \times E~, \quad {\rm {for}} \quad E > 0.5~ {\rm{GeV}}
 \end{equation}
  \begin{figure}[htb]
\centering
\includegraphics[width=\textwidth]{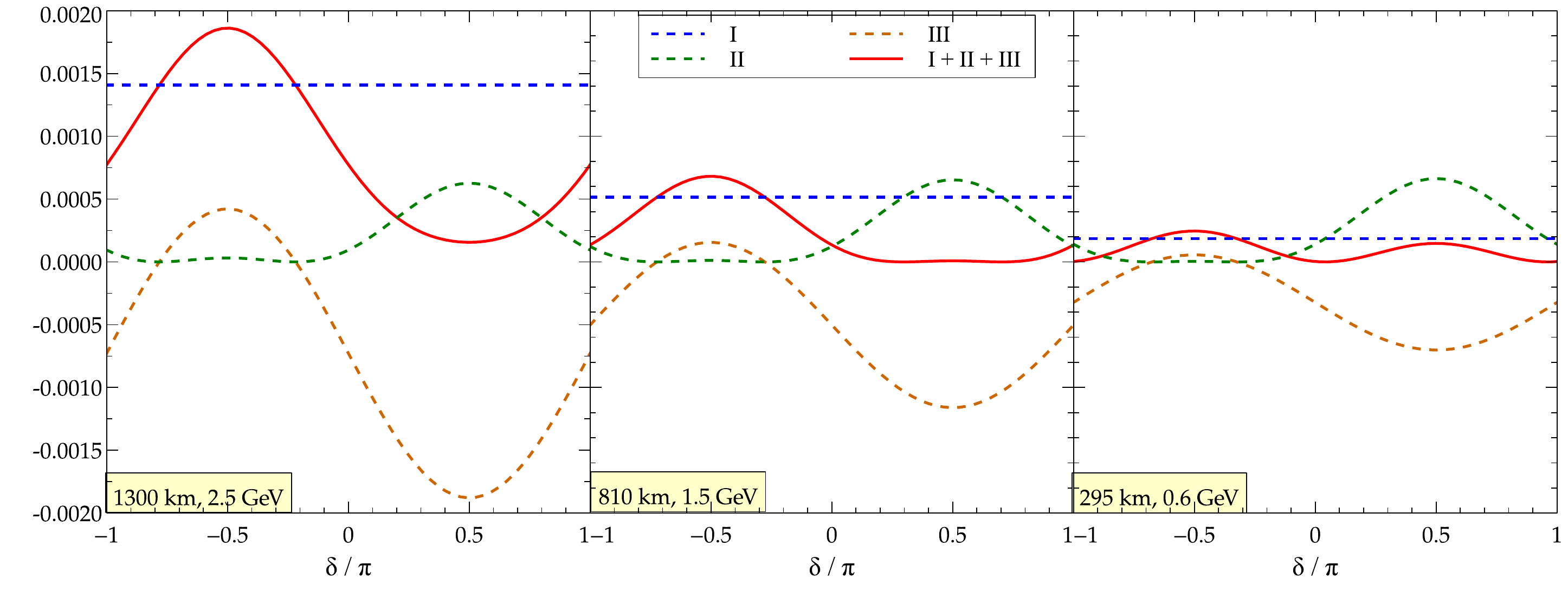}
\caption{\footnotesize{The numerator of the neutrino part in 
 Eq.~\ref{eq:chi_app} can be decomposed into 
three terms which are plotted separately (dashed) as well as combined (solid).
This demonstrates that the overall shape of the $\chi^2$ curves for mass ordering sensitivity
 depends upon the baseline and energy. 
 }}
\label{fig3}
\end{figure} 

 For  the disappearance channel, $P_{\mu e} $ is to be replaced by $P_{\mu \mu}$ and $\sigma_{\nu_e} \to \sigma_{\nu_\mu}$.  Note that 
 $\sigma_{\nu_\mu } \sim \sigma_{\nu_e}$ for the considered energy range~\cite{Messier:1999kj}.
  For antineutrinos, $\nu_\mu \to \bar\nu_\mu$ and $\nu_e \to \bar\nu_e$ and $P_{\mu e} \to \bar P _{\mu e}$. 
   
   %
%-------------------------------------------

%-------------------------------------------

 The $\chi^2$ for the question of mass ordering considering 
  the appearance  channel is obtained by adding the neutrino ($\nu_\mu \to \nu_e$) 
 and antineutrino ($\bar \nu_\mu \to \bar\nu_e $) 
 contributions\footnote{Note that the factors such as $N_{target}$, $\Phi_{\nu_\mu}$
  ($\bar\Phi_{\nu_\mu}$) and $\sigma_{\nu_\mu/\nu_e}$ ($\sigma_{\bar\nu_\mu/\bar\nu_e}$) 
 will also be present, but they are independent of the 
 CP phase and hence omitted in the discussion 
 that follows.} in Eq.\ \ref{eq:eventrate}  which is cast in terms 
 of difference between the true and the test event rates. 
 The difference in true and test event sample
  can be related to corresponding probability differences between NO and IO at different fixed 
  values of energies as shown in Fig.~\ref{fig1}. For any
   true fixed value of $\delta$ in case of NO i.e. lying on the magenta solid curve, one can easily see from Fig.\ \ref{fig1} that 
   the difference between the true fixed NO and test IO (all values of $\delta$ on the 
 blue solid curve) is minimum when $\delta(\text{test}) = -\pi/2$ for $E=2.5$ GeV~\footnote{For true fixed IO, $\delta (\text{test}) = \pi/2$ for $E=2.5$ GeV leads to the minimum difference.}. Hence for this energy we can express the $\chi^{2}$ as given below:
 {{
\begin{eqnarray}
\chi^2_{NO,app} (\delta_{tr}) &=& \chi^2_{\nu_\mu \to \nu_e} + 
\chi^2_{\bar \nu_\mu \to \bar \nu_e } ~,
\nonumber \\ &&
\!\!\!\!\!\!\!\!\!\!\!\!\!\!\!\!\!\!\!\!\!\!\!\!\!\!\!\!\!\!\!\!\!\!\!\!\!\!\!
\sim    \min_{\delta_{te} \in [-\pi,\pi]} \left\{
\frac{[P^{NO}_{\mu e} (\delta_{tr}) - P ^{IO}_{\mu e} (\delta_{te})]^2}
{P^{NO}_{\mu e} (\delta_{tr})} \times
\Phi_{\nu_\mu} \sigma _{\nu_e}
+ \frac{[\bar P^{NO}_{\mu e} (\delta_{tr}) -\bar P ^{IO}_{\mu e} 
(\delta_{te})]^2}{\bar P^{NO}_{\mu e} (\delta_{tr})} \times \bar 
\Phi_{\nu_\mu}  \sigma_{\bar \nu_e}
\right\}~,
\nonumber\\
&&
\!\!\!\!\!\!\!\!\!\!\!\!\!\!\!\!\!\!\!\!\!\!\!\!\!\!\!\!\!\!\!\!\!\!\!\!\!\!\!
\sim \left\{
\frac{[P^{NO}_{\mu e} (\delta_{tr}) - P ^{IO}_{\mu e} (\delta_{te} \sim -\pi/2)]^2}
{P^{NO}_{\mu e} (\delta_{tr})} \times
\Phi_{\nu_\mu} \sigma _{\nu_e}
+ \frac{[\bar P^{NO}_{\mu e} (\delta_{tr}) -\bar P ^{IO}_{\mu e} (\delta_{te} \sim -\pi/2)]^2}
{\bar P^{NO}_{\mu e} (\delta_{tr})}
\bar 
\Phi_{\nu_\mu}  \sigma _{\bar\nu_e} 
\right\}~,
\nonumber\\
&&
\!\!\!\!\!\!\!\!\!\!\!\!\!\!\!\!\!\!\!\!\!\!\!\!\!\!\!\!\!\!\!\!\!\!\!\!\!\!\!
\sim  \bigg\{
\frac{[\overbrace{
(x^2 - \bar x^2)^2}^{\textrm{I}}  + \overbrace{4 y^2 (  x \sin \delta_{tr} + \bar x )^2}^{\textrm{II}} + \overbrace{\{-4y (x\sin \delta_{tr} +\bar x) 
(x^2 -\bar x^2)\}}^{\textrm{III}} ]}{x^2 + y^2 - 2 x y \sin \delta_{tr}} \times 
\Phi_{\nu_\mu}  \sigma_{ \nu_e}  + \bar \nu~ {\textrm{term} }
 \bigg\}~.\nonumber\\
\label{eq:chi_app}
\end{eqnarray}
}}

Since the antineutrino cross-section and flux is smaller than that of neutrinos, 
the gross behaviour of $\chi^2$ plots can be understood by looking at the neutrino 
contribution in  Eq.~\ref{eq:chi_app}. The numerator contains three terms which are plotted in 
Fig.~\ref{fig3}. For the case of \dune, the first term $(x^2-\bar x^2)^2$ is a measure of matter 
effects and is independent of $\delta$ (shown as blue 
dashed line). The second term ($4 y^2 (x \sin \delta_{tr} + \bar x )^2$) is shown as 
green dashed curve which peaks at $\delta_{tr} \sim \pi/2$ and is flat elsewhere. 
 The third term ($- 4y (x\sin \delta_{tr} +\bar x ) (x^2 -\bar x^2) ]$) is shown as brown dashed curve 
 which has a peak at $\delta_{tr} \sim -\pi/2$ and a dip at $\delta_{tr} \sim \pi/2$. 
 Both the second and third terms  are $\delta-$dependent. The overall numerator is shown as red solid 
 curve which has a peak at $\delta \sim -\pi/2$ for NO.
For shorter baselines, \ttok\, and \nova, the contributions from the three terms at the respective peak energies are 
different  which lead to an overall feature of a primary peak at $\delta \sim -\pi/2$ and a secondary peak around $
\delta_{tr} \sim \pi/2$ for NO which was absent in case of \dune\,  and which can be attributed to the contribution coming from the third term  at different baselines. {{This simple decomposition of the net $\chi^2$ into
 terms appearing in the probability serves as a  useful guide to the expected shape of the $\chi^2$ curves and contrasting different baselines.   
To the best of our knowledge, this {\it{baseline-dependent characteristic shape of the $\chi^2$}} for mass ordering sensitivity has not been deduced from a simple probability level discussion in literature so far.}}

 % 
%%%%%%%%%%%%%%%%%%%%%%%%%%%%%%
For the true IO (blue solid curve) and test NO (magenta solid curve) in Fig.~2, we have $\delta_{te} 
\sim \pi/2$ (selected by marginalisation) for 
$E=2.5$ GeV,
{{
\begin{eqnarray}
\chi^2_{IO,app} (\delta_{tr}) &=& \chi^2_{\nu_\mu \to \nu_e} + \chi^2
_{\bar \nu_\mu \to \bar \nu_e } ~,
\nonumber \\
&&
\!\!\!\!\!\!\!\!\!\!\!\!\!\!\!\!\!\!\!\!\!\!\!\!\!\!\!\!\!\!\!\!\!\!\!\!\!\!\!
\sim 
 \min_{\delta_{te} \in [-\pi,\pi]} \left\{
\frac{[P^{IO}_{\mu e} (\delta_{tr}) - P ^{NO}_{\mu e} (\delta_{te})]^2}{P^{
IO}_{\mu e} (\delta_{tr})}  \times
\Phi_{\nu_\mu} \sigma _{\nu_e}
+ \frac{[\bar P^{IO}_{\mu e} (\delta_{tr}) -\bar P ^{NO}_{\mu e} 
(\delta_{te})]^2}{\bar P^{IO}_{\mu e} (\delta_{tr}) } 
 \times
\bar\Phi_{\nu_\mu} \sigma _{\bar\nu_e}
\right\}~,
\nonumber\\
&&
\!\!\!\!\!\!\!\!\!\!\!\!\!\!\!\!\!\!\!\!\!\!\!\!\!\!\!\!\!\!\!\!\!\!\!\!\!\!\!
\sim   \bigg\{
\frac{[(x^2 - \bar x^2)^2  + 4 y^2 ( \bar x \sin \delta_{tr} - x )^2 +
 4y (\bar x\sin \delta_{tr} - x ) 
(x^2 -\bar x^2) ]}{\bar x^2 + y^2 - 2 \bar x y \sin \delta_{tr}} 
\times  
\Phi_{\nu_\mu}  \sigma_{ \nu_e}
 + \bar \nu~ {\textrm{term} }
 \bigg\}~.\nonumber\\
\label{eq:chi_app_io}
\end{eqnarray}
}}
%
%~~~~~~~~~~~~~~~~~~~~~~~~~~~~~~~~~~~~~~~~~~~~~~~~~~~~~~~~~~~~~~~~~~~~~~~~~~~~~~~~~~~~~~~~~~~~~~~~~~~~~~~%
\section{Results}
\label{results}
%-------------------------------------------
\subsection{Impact of individual  NSI terms on mass ordering sensitivity at DUNE}
\label{results_b}
%-------------------------------------------

\begin{figure}[h!]
\centering
 \subfloat[Impact of $\varepsilon_{e\mu}$\label{em}]{%
      \includegraphics[width=0.85\textwidth]{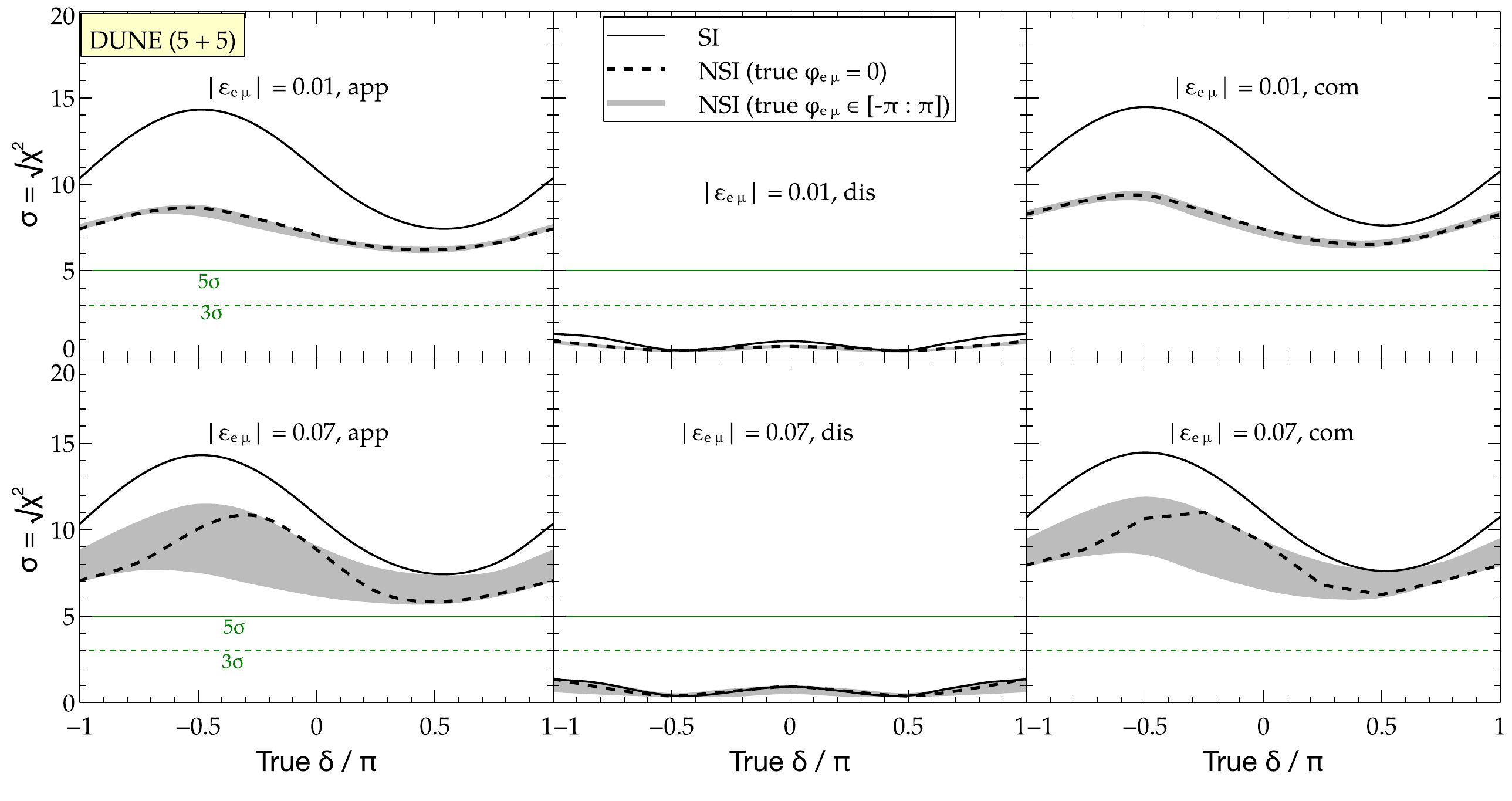}
    }
    \hfill
    \subfloat[Impact of $\varepsilon_{e\tau}$\label{et}]{%
      \includegraphics[width=0.85\textwidth]{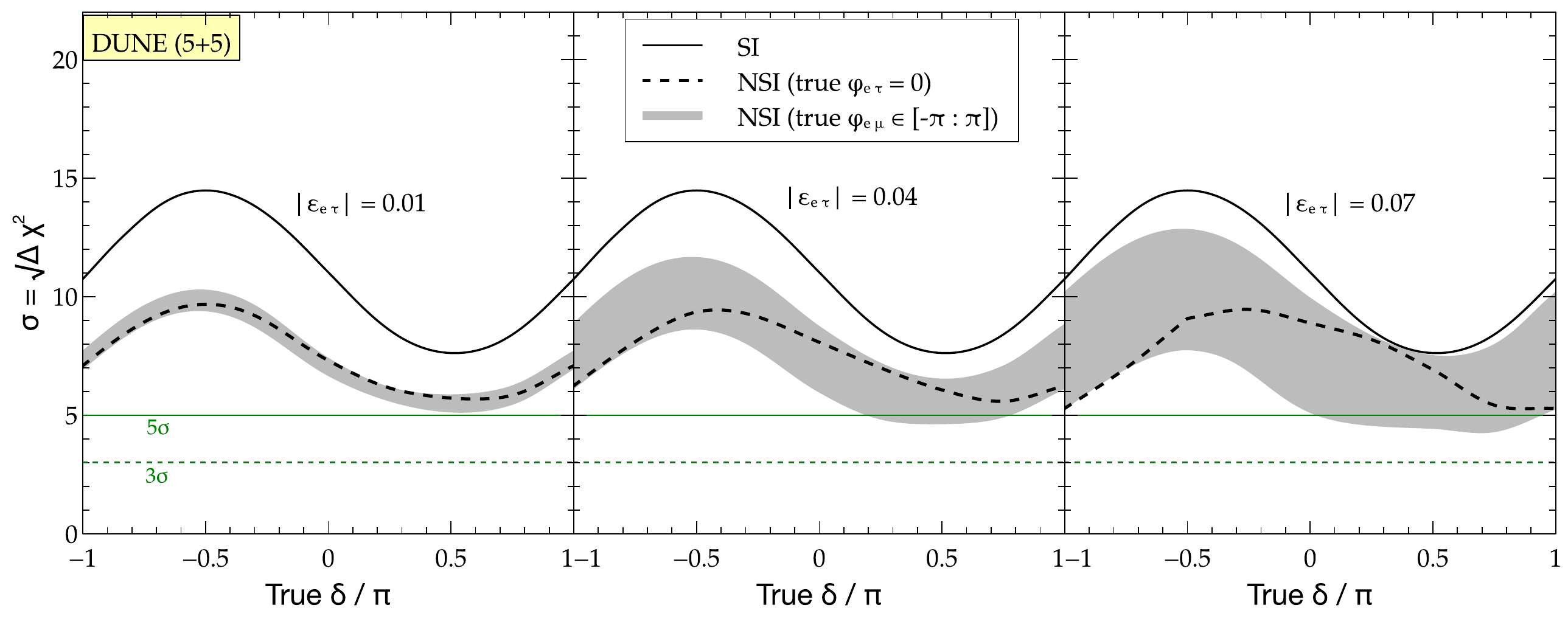}
      }
       \hfill
    \subfloat[Impact of $\varepsilon_{ee}, \varepsilon_{\mu\mu}, \varepsilon_{\tau\tau}$ 
    \label{eemmtt}]{%
      \includegraphics[width=0.85\textwidth]{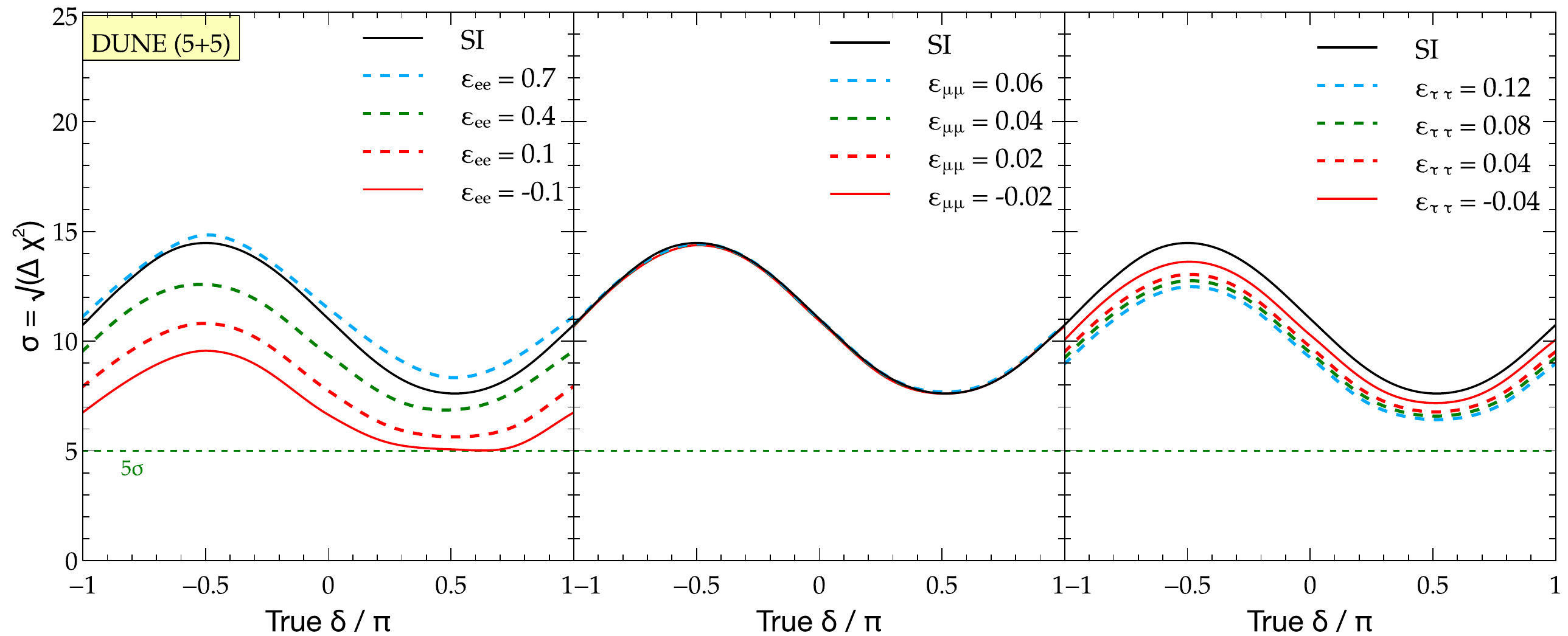}
    }
\caption{\footnotesize{{Impact of individual NSI terms on the mass ordering sensitivity. The top panel is for $\varepsilon_{e\mu}$, the middle panel is for $\varepsilon_{e\tau}$ and the last panel is for the three 
diagonal NSI parameters.}}
 }\label{fig:ind}
\end{figure}

In order to clearly understand the  impact of the NSI terms,
 we first consider only  one parameter non-zero at a time and discuss the role of appearance  ($\nu_\mu \to \nu_e$)  as well as the  disappearance  ($\nu_\mu \to \nu_\mu$) channels in addressing the question of ordering. 

 Before we describe the impact of a particular NSI parameter (i.e. $\varepsilon_{e\mu}$) 
 we would like to point out that there are two effects responsible for altering the value 
 of the $\chi^2$ which compete with each other as was first mentioned in Ref.~\cite{Masud:2016bvp} :
 \begin{description}
{\sl{\item (a) Decrease in the $\chi^2$ due to additional test values}} - NSI introduces more number of parameters 
 in the sensitivity analysis. If marginalization is carried out over more number of test 
 parameters (the difference from Ref.~\cite{Masud:2016bvp} is that
  $\varphi_{\alpha\beta}$ now takes more number of test values in the full range $ [-\pi,\pi]$ as opposed to $0,\pi$ only in the CP sensitivity case), it results in a decreased value of $\chi^2$. 
  This is purely a statistical effect. 
 \item 
{\sl{(b) Increase in the $\chi^2$ due to additional true values}} - For the mass ordering sensitivity, all the NSI 
 parameters (diagonal and off-diagonal ones) would serve as additional true values. 
In case of off-diagonal parameters, the variation over the NSI phases ($\varphi_{e\mu}$, $\varphi_{e\tau}$) tends to broaden the grey 
band provided the moduli ($|\varepsilon_{e\mu}|, |\varepsilon_{e \tau}|$) of the relevant NSI term is 
large. 
For the diagonal parameters (e.g.: $\varepsilon_{ee}$), if the value of the parameter is large enough 
(inducing more matter dependence which was indicated in Fig.~\ref{fig0}\footnote{The separation 
between NO and IO bands is visibly more pronounced (diminished) than the SI case when $\varepsilon_{ee}=
+0.7$ ($-0.7$) for $P_{\mu e}$.}), one can get a contribution to
 the $\chi^2$ which can be larger than SI case.
\end{description}

Let us now describe the impact of individual NSI parameters on the 
 mass ordering sensitivity expected for \dune\, as shown in Fig.~\ref{fig:ind}.
 The black solid curve depicts the case of SI while 
the black dashed curve is for NSI with zero NSI phases.  The shape of the black solid curve for the appearance channel for the baseline of \dune\, follows from the discussion in Sec.~\ref{framework_c} and Fig.~\ref{fig3}.
The impact of true non-zero NSI phases 
can be seen as grey bands for the choice of moduli of the NSI terms as mentioned in the legend.

For the case of $\varepsilon_{e\mu}$, the mass ordering sensitivity using both appearance and disappearance channels is shown (separately as well as combined) in  Fig.~\ref{fig:ind}(a). It can be seen that the disappearance channel contributes very little\footnote{This can be 
attributed to the small matter dependence in $P_{\mu\mu}$ at the baselines considered~\cite{Gandhi:2004bj}
(see the previous section also).} and the appearance channel  is the main contributor to the total $\chi^2$. 
If $|\varepsilon_{e\mu}| = 0.01$ as shown in the top row,  the overall $\chi^2$ 
decreases due to effect (a) for all values of $\delta$ while (b) is insignificant. 
However, for somewhat larger value such as $|\varepsilon_{e\mu}| = 0.07$ as shown 
in the second row of Fig.~\ref{fig:ind}(a), we note that the net $\chi^2$ again
 decreases for all values of $\delta$ but the non-negligible impact of effect (b) can be
  inferred from the broadening of the grey bands. 
  
  The impact of $\varepsilon_{e\tau}$  is shown 
  for three different values for the combined (appearance  + disappearance) case in  Fig.~\ref{fig:ind}(b). 
  The effect is  similar to that of $\varepsilon_{e\mu}$ described above but 
  the widening of grey bands due to the effect of relevant true phase variation ($\varphi_{e\tau}$)  is some what more   pronounced in comparison to Fig.~\ref{fig:ind}(a). The widening of grey bands depends on the value of $|\varepsilon_{e\tau}|$. As we go from smaller to larger value of  $|\varepsilon_{e\tau}|$, we see that NSI can drastically alter the $\chi^2$  at all values of $\delta$.  However, the $\chi^2$ always stays below the SI expectation for the values considered.

%-------------------------------------------
\begin{figure}[ht!]
\centering
\includegraphics[width=0.9\textwidth]
{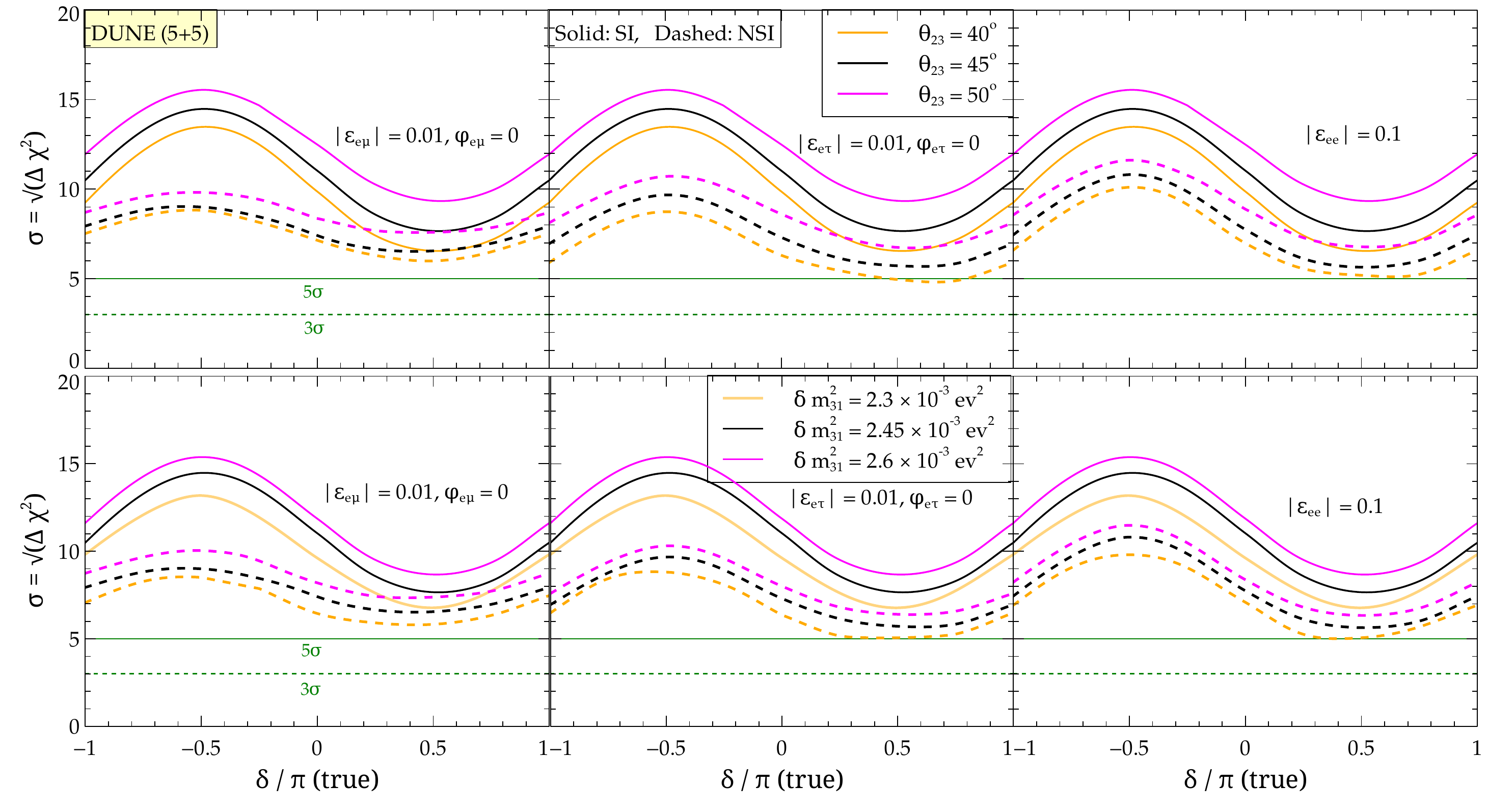}
\caption{\footnotesize{Dependence of the mass ordering sensitivity  on the value of
 $\theta_{23}$ and $\delta m^2_{31}$.
 }}
\label{fig:chisq_th23_dm31}
\end{figure}
%-------------------------------------------

The impact of all the three diagonal NC NSI parameters is shown in 
Fig.~\ref{fig:ind}(c) for positive and negative values of the relevant NSI parameters.
In general we note that for diagonal NSI parameters, the effect (a) is somewhat reduced 
due to the fact that there are no phases i.e. lesser parameters being marginalised over. This easily allows effect (b) to overtake (a) if the value of NSI parameter is large enough. 
The effect of $\varepsilon_{ee}$ is shown on the left for positive and negative values of $\varepsilon_{ee}$. For 
$\varepsilon_{ee} = 0.7$, the effect (b) overtakes (a). The case of negative NSI parameter is also shown for comparison.  For  $\varepsilon_{\mu\mu}$, there is hardly any impact since $\varepsilon_{\mu\mu}$ is constrained very well. For $\varepsilon_
{\tau\tau}$, we note that effect (a) dominates the sensitivity plots and the general
 trend is opposite to that seen for $\varepsilon_{ee}$.

%-------------------------------------------
\subsection{Dependence on  the value of $\theta_{23}$ and $\delta m^2_{31}$}
\label{results_c}
%-------------------------------------------

We depict the dependence of  the mass ordering sensitivity on the choice of true values of $\theta_{23}$ and $\delta m^2_{31}$ in Fig.~\ref{fig:chisq_th23_dm31}. The top row shows the impact of  $\theta_{23}$, we note that the SI curves are consistent with Ref.~\cite{Acciarri:2015uup} i.e. larger the 
value of $\theta_{23}$, higher will be the sensitivity to neutrino mass ordering. In presence of NSI, the sensitivity goes down but still remains above $5\sigma$ for most values of $\delta$ for the choice of parameters considered in the figure in accordance with the discussion in the preceeding section. 
Again as far as $\theta_{23}$ dependence is concerned, the trend is similar to that in case of SI.

The bottom row shows the impact of choice of value of $\delta m^2_{31}$. Here again, the SI curves are consistent with Ref.~\cite{Acciarri:2015uup} i.e., a larger value of $\delta m^2_{31}$ aids in determination of the neutrino mass ordering. We see a similar trend in case of NSI among the values of $\delta m^2_{31}$ considered. The overall  mass ordering sensitivity is lower than the SI case (while still above $5\sigma$) as expected from the discussion in the previous section. 

Also, in both the panels for SI and NSI, we see a trend that is opposite to that seen in the 
 CP sensitivity plots~\cite{Masud:2016bvp}.

%-------------------------------------------
\subsection{Comparison with other long baseline experiments}
\label{results_d}
%-------------------------------------------

\begin{table}[htb]
\centering
\begin{tabular}{|l | c c | c c  |}
\hline
Experiment & \multicolumn{2}{c |}{Appearance channel} & \multicolumn{2}{c 
|}{Disappearance channel} \\
                                         \cline{2-5}
                                        & $\nu_\mu \to \nu_e$ 
                                        & 
                                         $\bar\nu_\mu \to \bar\nu_e$ & 
                                         $\nu_\mu \to \nu_\mu$ &
                                          $\bar \nu_\mu \to \bar\nu_\mu$                                                                                      
                                        \\ 
\hline
&&&&\\
{{\dune}} &&&&\\
NO, $\delta=-\pi/2$ & 1610/1779 & 229/214 & 11431/11418 & 7841/7840 \\
NO, $\delta=0$ & 1350/1803 & 292/257 & 11401/11313 & 7805/7826\\
NO, $\delta=\pi/2$ & 1028/1182 & 309/279 & 11431/11417 & 7841/7840\\
IO, $\delta=-\pi/2$ & 861/740 & 357/392 & 11052/11036 & 7611/7609 \\
IO, $\delta=0$ & 610/383 & 416/529 & 11110/11188 & 7631/7612 \\
IO, $\delta=\pi/2$ & 468/399 & 499/540 & 11052/11036 & 7611/7609 \\
 \hline
{{\nova}} &&&&\\
NO, $\delta=-\pi/2$ & 90/95 & 17/15 & 142/143 & 46/47 \\
NO, $\delta=0$ & 78/93 & 25/20 & 141/140 & 45/46\\
NO, $\delta=\pi/2$ & 57/62 & 30/27 & 142/143 & 46/47\\
IO, $\delta=-\pi/2$ & 63/58 & 27/29 & 130/129 & 41/41\\
IO, $\delta=0$ & 46/36 & 33/40 & 131/132 & 42/41\\
IO, $\delta=\pi/2$ & 36/33 & 42/45 & 130/129 & 41/41 \\
 \hline
{{\ttok}} &&&&\\
NO, $\delta=-\pi/2$ & 127/130 & 20/19 & 372/371 & 130/129 \\
NO, $\delta=0$ & 111/119 & 29/27 & 367/365 & 127/128\\
NO, $\delta=\pi/2$ & 80/82 & 33/32 & 372/371 & 130/129\\
IO, $\delta=-\pi/2$ & 113/110 & 23/24 & 335/336 & 116/117 \\
IO, $\delta=0$ & 83/77 & 28/31 & 340/341 & 119/118 \\
IO, $\delta=\pi/2$ & 68/66 & 37/38 & 335/336 & 116/117 \\
 \hline
\end{tabular}
\caption{\label{tab:events} Total number of signal events (SI/NSI) summed over all energy bins for each experiment using the oscillation parameters given in Table 1. 
For NSI, we show the collective case when the NSI parameters  $|\eem|=0.07, |\eet|=0.07, |\eee|=0.7$, $\varphi_{e\mu}=0$ and $\varphi_{e\tau}=0$  are considered.}
\end{table}
%-------------------------------------------

 Below we give a very brief description\footnote{for more details, see 
\cite{Masud:2016bvp}} of the ongoing and future experiments that would be sensitive to the question 
 of neutrino mass ordering using the two oscillation channels considered in the present work.  

\begin{description}
\item \ttok : 
The \ttok\, (Tokai to Kamioka) experiment  aims neutrinos from the Tokai site to the SK detector
 295 km away (see~\cite{TheT2KCollaboration01042015}). The peak energy is $0.6$ GeV. We consider a
  runtime of $3\nu+3\bar\nu$ and
 a 22.5 kton Water Cherenkov 
 (WC) detector.

\item \nova :
The NuMI Off-axis $\nu_e$ Appearance (\nova) experiment has a baseline of 810 km, and the detector is exposed to an off-axis 
(0.8 degrees) neutrino beam produced from 120 GeV protons at Fermilab (see \cite{nova, Adamson:2016tbq}). We consider a
  runtime of $3\nu+3\bar\nu$ and a 14 kton Totally Active Scintillator Detector (TASD). 

\item \dune : The \dune\, experiment has a baseline of 1300 km and the $\nu_\mu$ beam peaks at $2.5$ GeV (see 
\cite{Acciarri:2015uup}). 
 We consider a $35$ kton fiducial liquid argon detector placed at 1300 km 
 from the source, on-axis with respect to the beam direction. We consider a runtime of $5 \nu + 5 \bar\nu$ years.

 \end{description}
\begin{figure}[ht!]
\centering
\includegraphics[width=0.9\textwidth]
{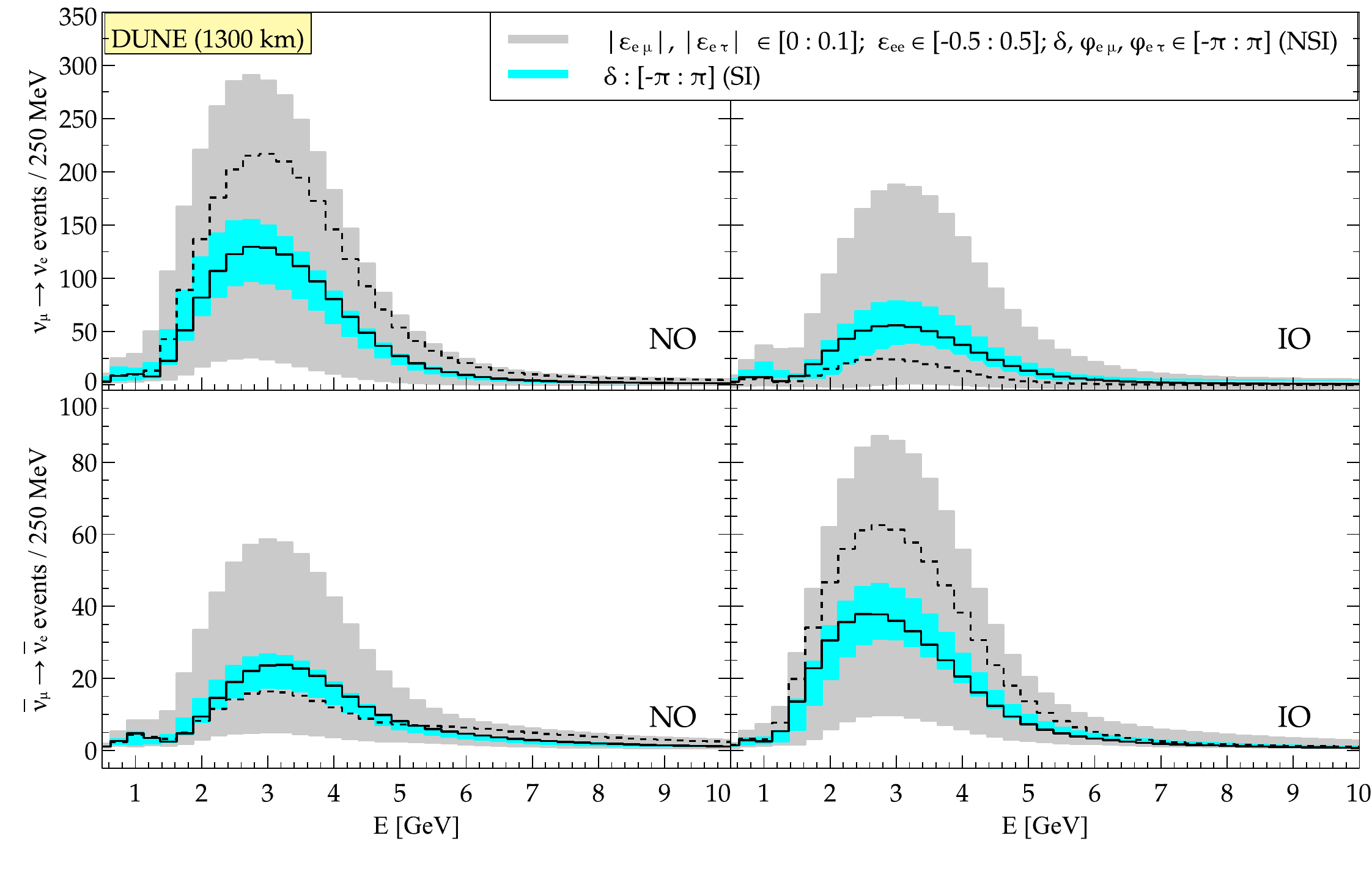}
\caption{\footnotesize{Variation of the neutrino and the antineutrino event rate histograms (for the appearance channel) are shown at DUNE for both SI (cyan) and the NSI (grey) scenario due to the variation in the relevant parameters as indicated in the legend. The dashed (solid) black curve indicates the case when all the relevant NSI (SI) phases are zero. 5 yrs. of neutrino and 5 yrs. of antineutrino runs were considered. The left (right) panel depicts the case of normal (inverted) mass ordering.  
 }}
\label{fig:e1}
\end{figure}
 
 In  Table~\ref{tab:events}, we give
  the energy integrated events\footnote{The energy range for the various experiments
  is mentioned in Sec.~\ref{framework_c}} for the mentioned experiments and channels  
   for neutrinos as well as antineutrinos for the two possible orderings. 
  As expected, the events for the disappearance channel are much larger than  the 
   appearance channel. 
   The event rates for \ttok\, and \nova\, are comparable due to the 
    larger detector size of \ttok\, which  compensates for the 
    shorter baseline when compared with \nova\,  with a longer baseline.  
    The event rates are higher for \dune\, due the  bigger detector size 
    in comparison to  \nova\, even though the baseline for \dune\, is larger than \nova.

 Let us look at the impact of NSI at the level of event rates for the choice of parameters considered. 
%
%-------------------------------------------

%-------------------------------------------
We show the event rate histograms at DUNE for the $\nu_{e}$ appearance (Fig.~\ref{fig:e1}) and the $\nu_{\mu}$ disappearance (Fig.\ \ref{fig:e2}) channels. The cyan bands show the  variation of events due to the full variation of the standard phase $\delta$ in the SI scenario. The grey bands indicate the variation in the NSI event spectra when the dominant NSI parameters ($\varepsilon_{e \mu}$, $\varepsilon_{e \tau}$ and $\epsilon_{ee}$) have been varied in addition to the variation of standard $\delta$ within the ranges indicated in the legends of the plots. We can make the following observations from Figs.\ \ref{fig:e1} and \ref{fig:e2}:

%-------------------------------------------
\begin{figure}[ht!]
\centering
\includegraphics[width=0.9\textwidth]
{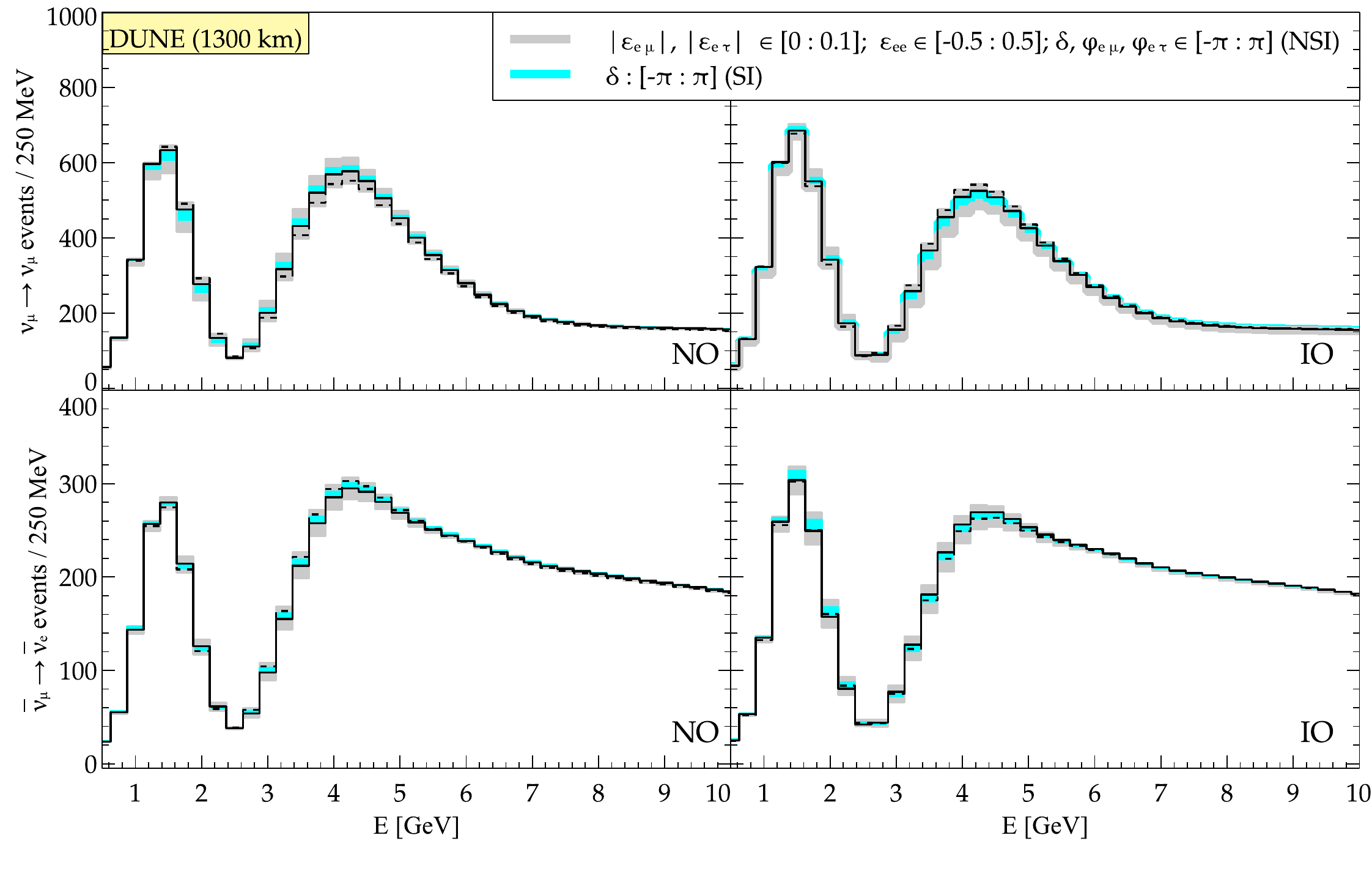}
\caption{\footnotesize{Similar to Fig.\ \ref{fig:e1}, but for the disappearance channel.
 }}
\label{fig:e2}
\end{figure}
%-------------------------------------------

\begin{itemize}
 \item In Fig.\ \ref{fig:e1}, the grey bands (NSI) span much further on both sides of the cyan (SI) bands and this feature is most prominent around the peak ($\sim 2-5$ GeV). This implies that a significant surfeit or dearth of appearance events around the peak of the energy is a signature of the presence of new physics. On the other hand, Fig.\ \ref{fig:e2} shows that the effect of NSI for the disappearance channel is very small and is manifested as the much thinner grey bands. This is because the matter effect in the disappearance channel is very small 
 \footnote{The matter effect in the $P_{\mu\mu}$ channel becomes significant only when the baseline 
 chosen is very long~\cite{Gandhi:2004bj}.}. Consequently the relative change in the disappearance event rate due to the presence of NSI is much smaller than that of the appearance events shown in Fig.\ \ref{fig:e1}. 
 
 \item We note that both the NSI and the SI event spectra in Fig.\ \ref{fig:e1} fall very close to zero around energies $\gtrsim 6.5$ GeV onwards, because of the rapidly falling flux as well as the probability $P_{\mu e}$ (or $\bar{P}_{\mu e}$). But, for the disappearance channel in Fig.\ \ref{fig:e2}, even if the flux falls rapidly, the tail of the event spectra, unlike Fig.\ \ref{fig:e1}, is significantly high and falls much slowly. This is due to the large value of $P_{\mu\mu}$ (or $\bar{P}_{\mu\mu}$) even at energies $\sim 6-10$ GeV.
 
 \item The dashed black curve (NSI event spectra corresponding to zero phases) in Fig.\ \ref{fig:e1} is greater or less than the solid black curve (SI event spectra) depending on the mass ordering and  polarity. This trend is also similar for Fig.\ \ref{fig:e2}, although due to the much smaller effect of NSI, the dashed black curves are very close to the solid ones.
  
\end{itemize}

%-------------------------------------------
\begin{figure}[h!]
\centering
\includegraphics[width=.8\textwidth]
{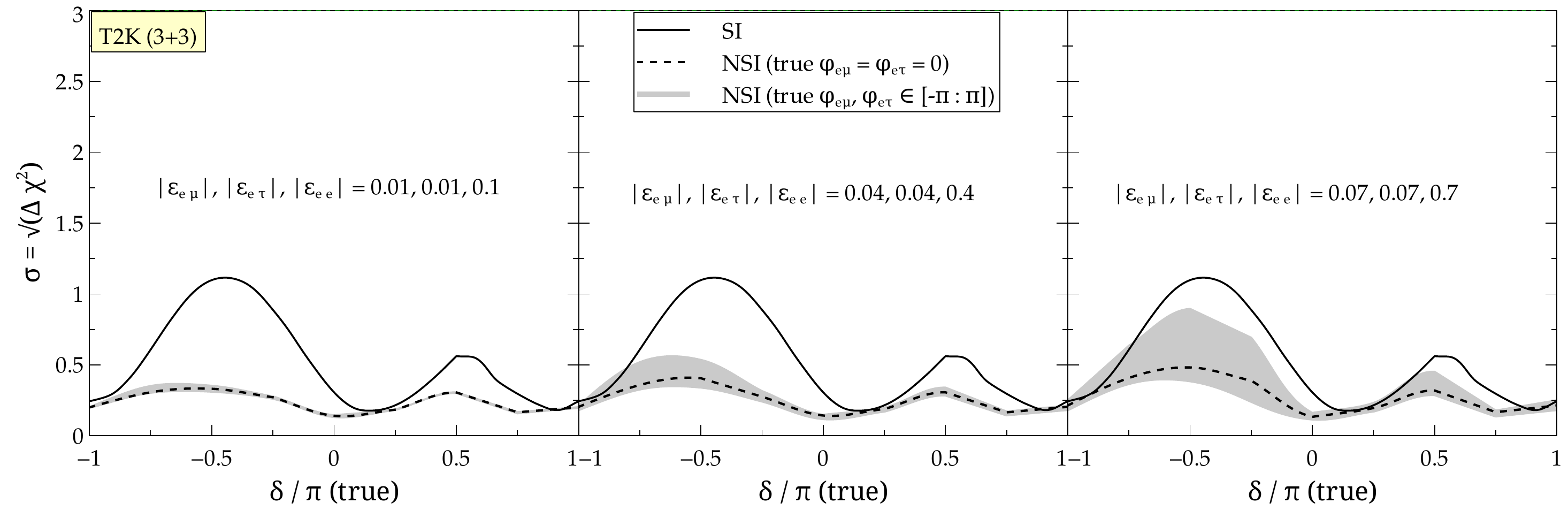}
\includegraphics[width=.8\textwidth]
{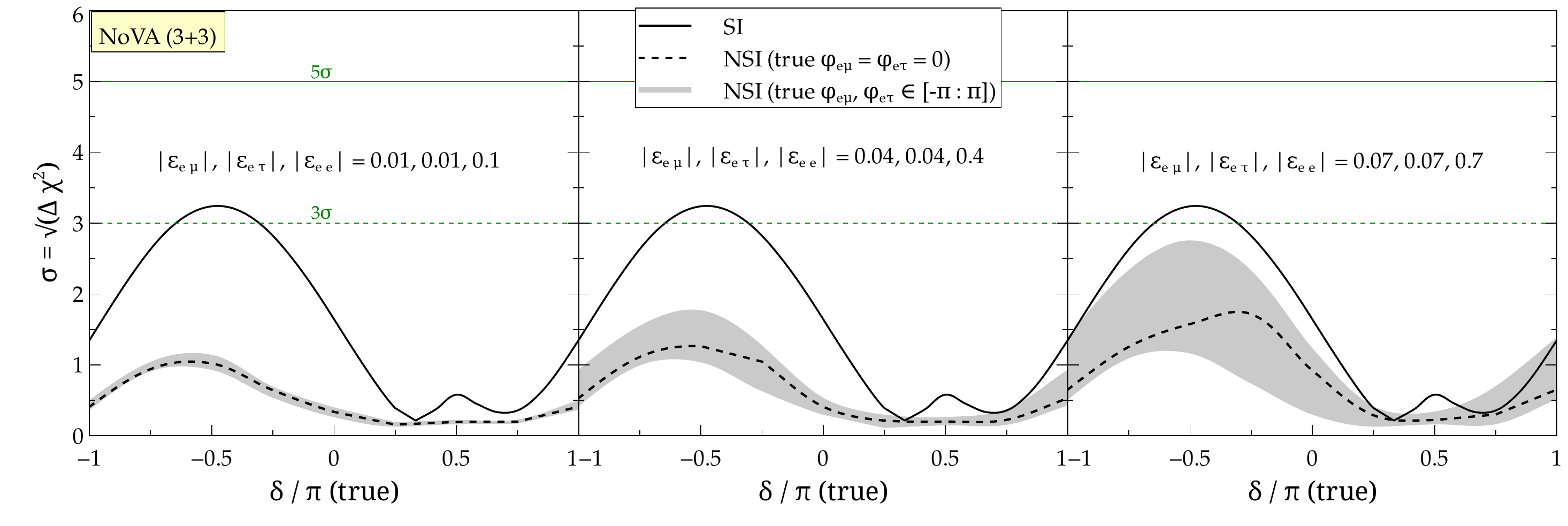}
\includegraphics[width=.8\textwidth]
{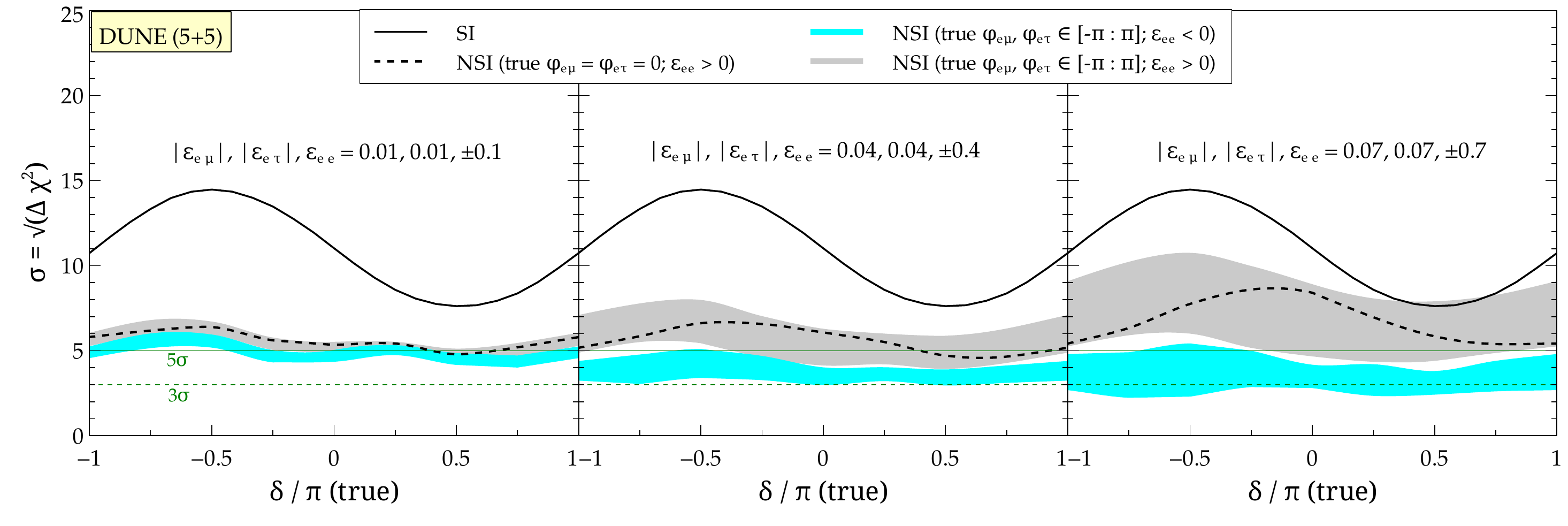}
\caption{\footnotesize{Mass ordering sensitivity at \ttok, \nova, and \dune\, 
 for collective NSI case and SI as a function of true $\delta$ (NO).  
 }}
\label{fig:mo1}
\end{figure}
%-------------------------------------------

%-------------------------------------------
\begin{figure}[htb]
\centering
\includegraphics[width=.9\textwidth]
{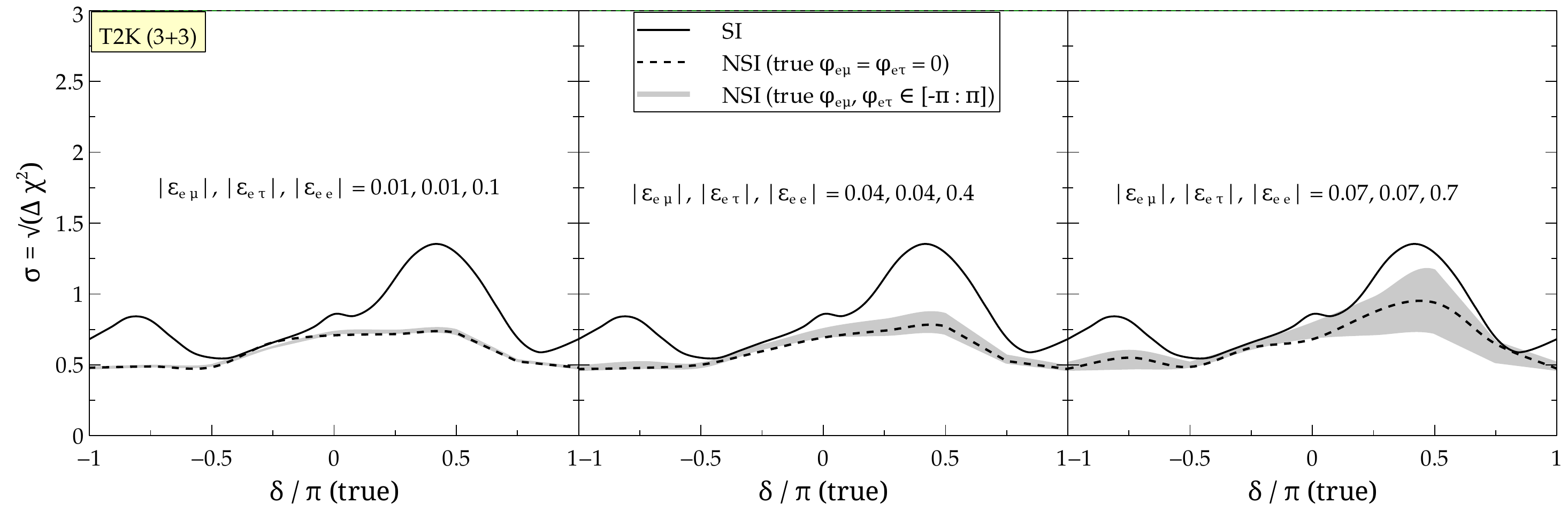}
\includegraphics[width=.9\textwidth]
{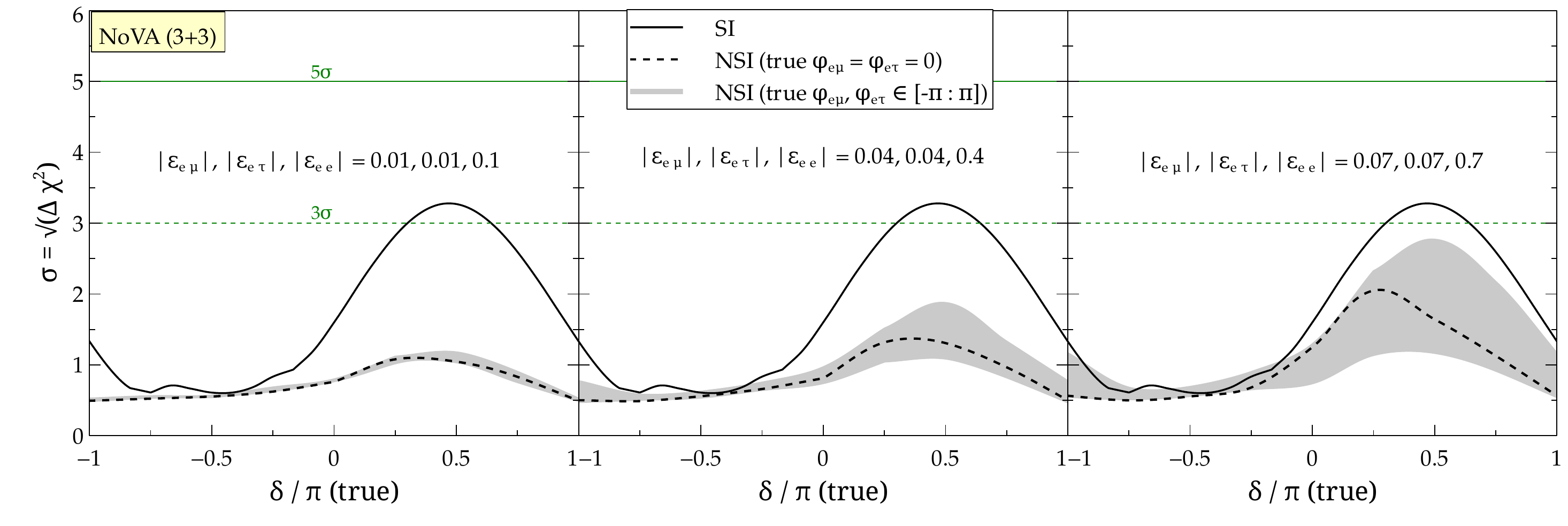}
\includegraphics[width=.9\textwidth]
{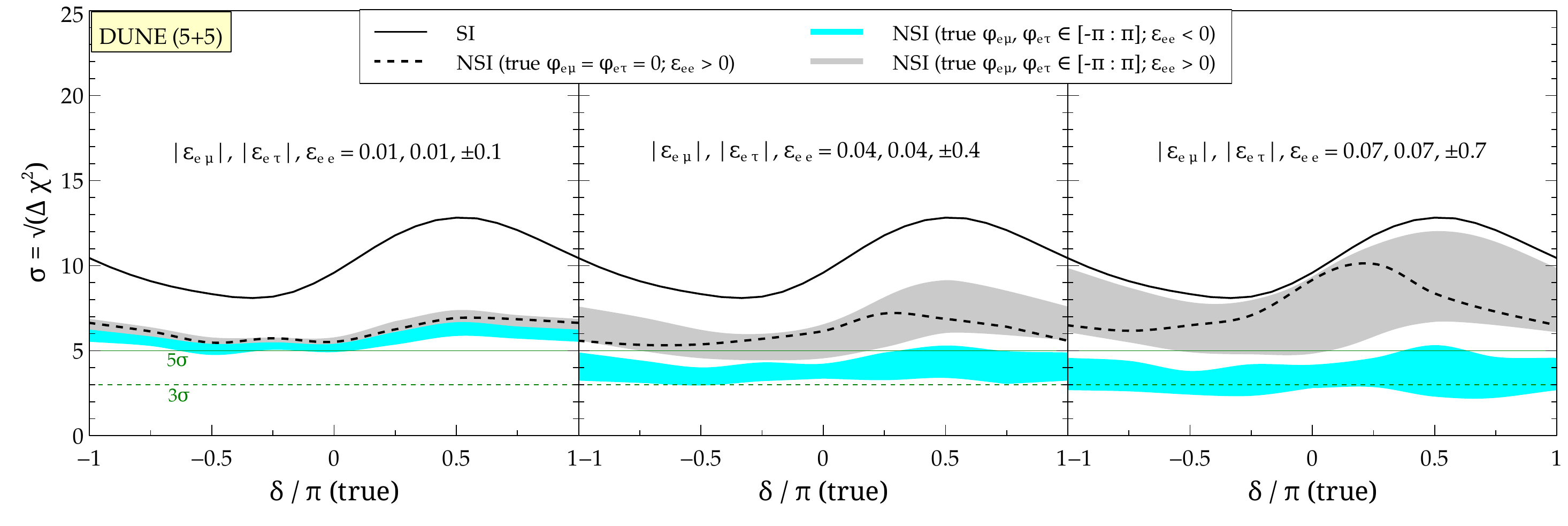}
\caption{\footnotesize{Mass ordering  sensitivity at \ttok, \nova, and \dune\, 
 for collective NSI case and SI as a function of true $\delta$ (IO). 
 }}
\label{fig:mo2}
\end{figure}
%
%---------------------------------------------

  The expected  mass ordering sensitivity offered by different experiments 
   is illustrated in Figs.~\ref{fig:mo1} and \ref{fig:mo2} for true NO and true IO respectively  considering the relevant NSI parameters
    collectively.

In Fig.~\ref{fig:mo1} (Fig.~\ref{fig:mo2}), we show a comparison of mass ordering sensitivity for the three 
experiments with and without NSI for the case of true NO (IO). We can immediately notice that the (solid black) curves for SI resemble 
 the characteristic shape based on the  statistical definition of $\chi^2$  described earlier (see Sec.~\ref{framework_c} and Fig.~\ref{fig3}). 
 In presence of NSI, the {{baseline-dependent characteristic shape of the $\chi^2$}} for mass 
ordering sensitivity is  spoiled depending upon the baseline and the size of the NSI term. 
This distortion in shape is expectedly more for the longer baselines considered. 
In addition, due to the 
effect (a) mentioned in subsection~\ref{results_b}, there are effects such as suppression in the value of
 $\chi^2$ for the values of NSI parameters  considered. 
 
For \ttok\, and \nova\,   individually, we note that the mass ordering sensitivity almost never reaches $3\sigma$ (it barely touches 
 $\sim 1.1\sigma$ (for \ttok) and $\sim 3.2\sigma$ (for \nova)) when SI is present (see Fig.~\ref{fig:mo1}, first and second row). 
  This means that these two current experiments considered in isolation 
  are not so much interesting as far as 
  mass ordering sensitivity is concerned. 
  This does not come as a surprise as \ttok\, and \nova\, are shorter baseline experiments (compared to 
  \dune) and hence do not give the best sensitivity to the neutrino mass ordering. 
 Even combining the data from \ttok\, and \nova\, would be a futile exercise for most values of $\delta$ for SI. In presence of NSI, the value of $\chi^2$ undergoes a suppression in general for all values of $\delta$.

  If we look at the mass ordering sensitivity expected from \dune\, (see Fig.~\ref{fig:mo1}, third row), we note that it is much above $5\sigma$ for all values of true  $\delta$ in presence of SI which  means that it is not needed to combine data from the \ttok\, and \nova\, to \dune. In presence of NSI, there is a suppression in the value of $\chi^2$ for all values of $\delta$. However for most values of $\delta$, it stays above $5\sigma$ if the  NSI parameter $\varepsilon_{ee}$ is positive. If $\varepsilon_{ee} < 0$, the cyan band shows that the $\chi^2$ can get further suppressed and can lie in the range $3-5 \sigma$ if the NSI terms are large enough. 
   
In Fig.~\ref{fig:mo2}, the mass ordering sensitivity is shown for the case of IO. The characteristic shapes of the SI curves can also be predicted from the discussion in Sec.~\ref{framework_c}. The impact of NSI is similar to that for the case of NO. 
 
\subsection{Degeneracy and the issue of mass ordering}
\label{deg}

%-------------------------------------------
\begin{figure}[ht!]
\centering
\includegraphics[width=0.9\textwidth]
{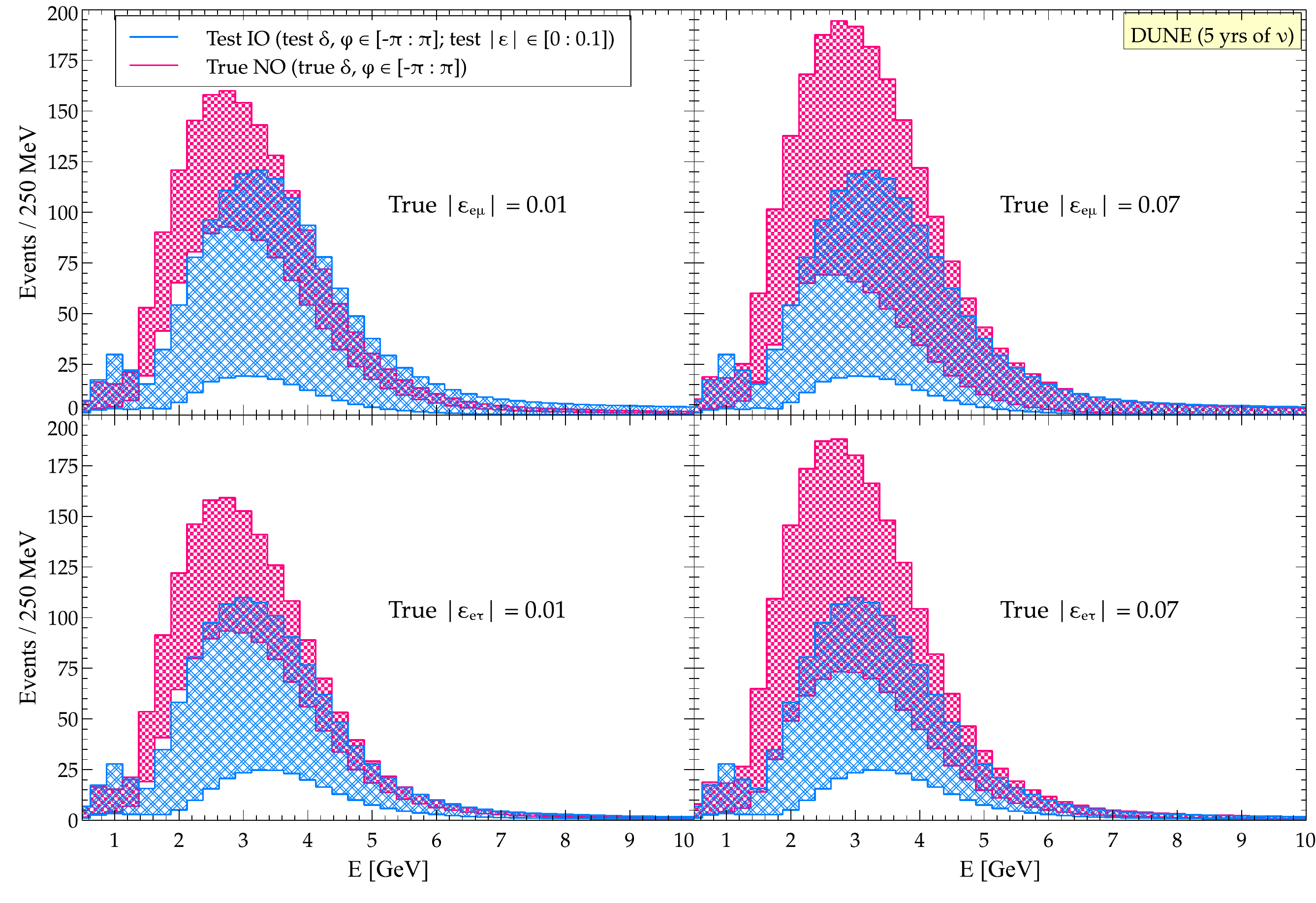}
\caption{\footnotesize{Variation of the neutrino event rate histograms for the appearance channel  at DUNE for the NSI scenario (taken one off-diagonal NSI parameter at a time) for NO (true) and IO (test) data set. The magenta (blue) shaded region corresponds to the NO (IO) variation due to the variation of the relevant parameters (see text for details) as shown in the legend.
 5 yrs. of neutrino  run was considered. The different panels are for different values of the true NSI parameter considered.  
 }}
\label{fig:e3}
\end{figure}
%-------------------------------------------

\begin{figure}[ht!]
\centering
\includegraphics[width=0.9\textwidth]
{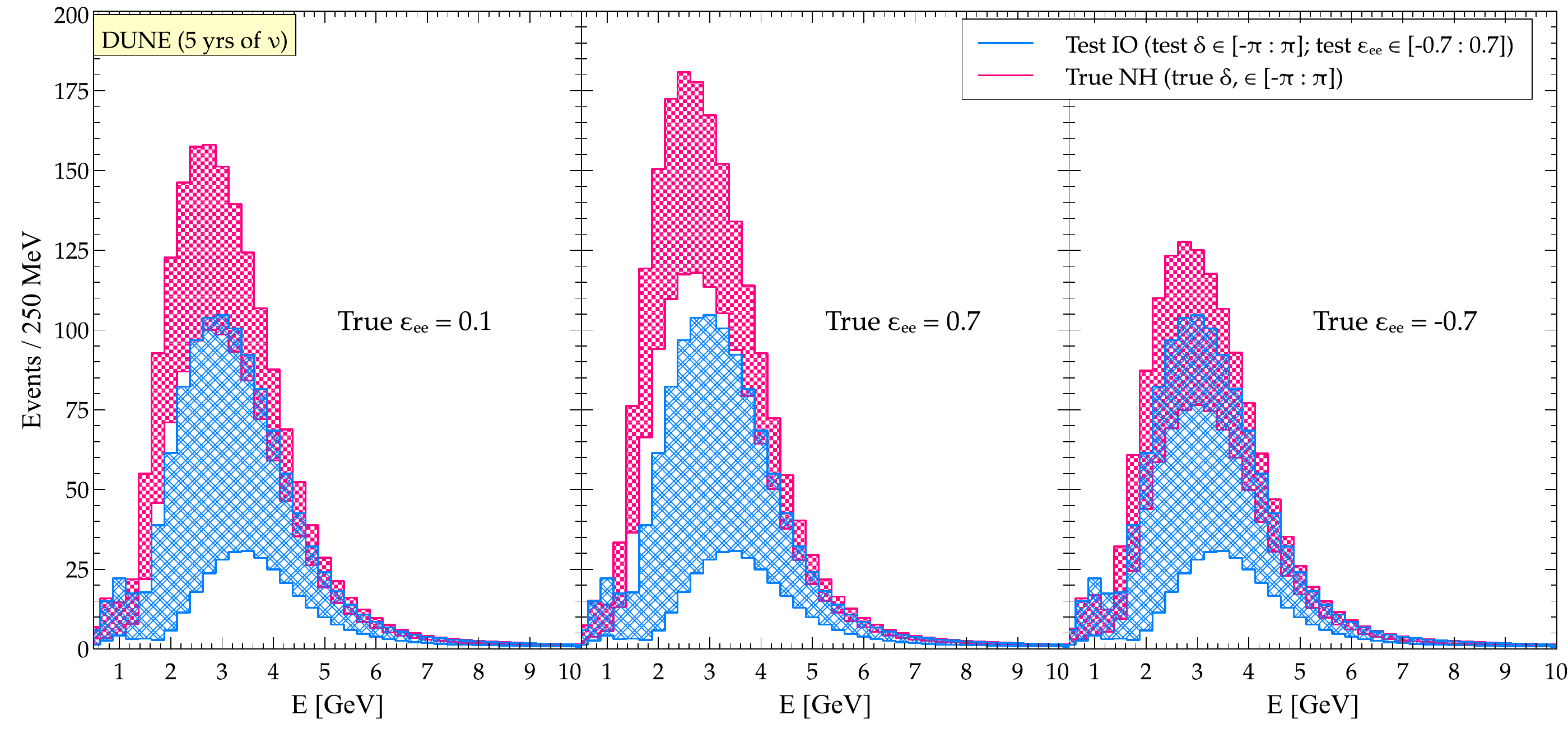}
\caption{\footnotesize{Same as Fig.~\ref{fig:e3} but for different diagonal NSI parameter, $\varepsilon_{ee}$.
 }}
\label{fig:e4}
\end{figure}

Let us analyse the information contained in Fig.\ \ref{fig0} in terms of  event rate histograms for the appearance 
channel at DUNE for the  off-diagonal and diagonal NSI parameters respectively (Figs.~\ref{fig:e3} and \ref{fig:e4}). 
These plots provide another perspective to understand the impact of NSI on the question of mass ordering. 
 The NO band in Fig.\ \ref{fig:e3}, (true data set corresponding to Fig.~\ref{fig:ind}) is obtained by fixing  the  NSI parameter 
(mentioned in the plot) is obtained by varying the relevant parameters 
$\delta$ and $\varphi_{e\mu}$ (or $\varphi_{e\tau}$) in each panel. The IO band (test data set) in  Fig.\ \ref{fig:e3},
is obtained by varying the relevant parameters 
$\delta$ and $\varphi_{e\mu}$ (or $\varphi_{e\tau}$) as well as $|\varepsilon_{e\mu}|$ (or $|\varepsilon_{e\tau}|$) in each panel. 
This shows the maximum possible variation in the event rates for NO and IO in Fig.~\ref{fig:e3}. 
 The NO ({true}) band is for a fixed value of the 
 $|\varepsilon_{e\mu}|$ (or $|\varepsilon_{e\tau}|$) and the IO ({test}) band corresponds to the case where 
  $|\varepsilon_{e\mu}|$ (or $|\varepsilon_{e\tau}|$) are also allowed to vary.

The $\Delta \chi^{2}$, as shown in Fig.\ \ref{fig:ind} and  explained in subsection \ref{framework_c}, is roughly a difference between the NO ({true}) and the IO ({test}) events for two fixed values of true $|\varepsilon_{e\mu}| = 0.01, 0.07$.  The possibility of distinction between NO and IO depends upon the overlap
 between NO and IO bands and increases as the overlap becomes small.  
 The NO bands in Fig.\ \ref{fig:e3} which are shown as magenta shaded regions correspond to
  fixed true values of  $|\varepsilon_{e\mu}|$ or $|\varepsilon_{e\tau}|$. 
  Clearly, as the modulus of the relevant parameter 
 increases,  the shaded region becomes wider. 
 The IO bands are shown as blue shaded regions and incorporate the entire range of  
 $|\varepsilon_{e\mu}|$ or $|\varepsilon_{e\tau}|$. Thus, as we go from left to right, these bands remain unchanged. As the true value of   $|\varepsilon_{e\mu}|$ or $|\varepsilon_{e\tau}|$ increases, the NO band widens  while the IO band remains unchanged. As a result, the overlap between NO and IO bands increases. This would imply that the sensitivity to mass ordering is impacted severely when relevant NSI parameter takes a larger value.
 This is also illustrated by means of sensitivity curves in Fig.~\ref{fig:ind} since the change in sensitivity is proportional to the magnitude of the NSI parameter considered.

For the diagonal NSI parameter ($\varepsilon_{ee}$), we can make similar observations  
from Fig.~\ref{fig:e4}. Note that in Fig.~\ref{fig:ind}c (left panel) a positive value of $\varepsilon_{ee}$ can enhance
 the sensitivity to mass ordering whereas a negative value of $\varepsilon_{ee}$ suppresses it. 
Similar conclusions are also indicated in Fig.~\ref{fig:e4} i.e. the NO and the IO bands are closer together 
for a negative $\varepsilon_{ee}$ and get further apart as $\varepsilon_{ee}$ increases in the 
positive direction.

%-------------------------------------------

%-------------------------------------------

%-------------------------------------------
\subsection{Optimal exposure for mass ordering discovery}
\label{results_e}
%-------------------------------------------

All the plots shown thus far were obtained by keeping the total exposure fixed  for a given experimental configuration. It is pertinent to ask  what would be the optimal exposure for a given experiment that would be required to decipher the neutrino mass ordering.  

We can define a useful quantity called the {\bf{mass ordering fraction $f ^{MO}(\sigma > 5)$}} that can be used to quantify the ability of a given experiment to determine the neutrino mass ordering.  $f ^{MO}(\sigma > 5)$
  refers to the fraction of true $\delta $ values for which 
mass ordering can be determined above a  particular significance (here, $5\sigma$).  Being a fraction, 
$f^{MO} (\sigma > 5)$ naturally lies between $0$ and $1$.
%  

%-------------------------------------------

\begin{figure}[htb]
\centering
\includegraphics[width=\textwidth]{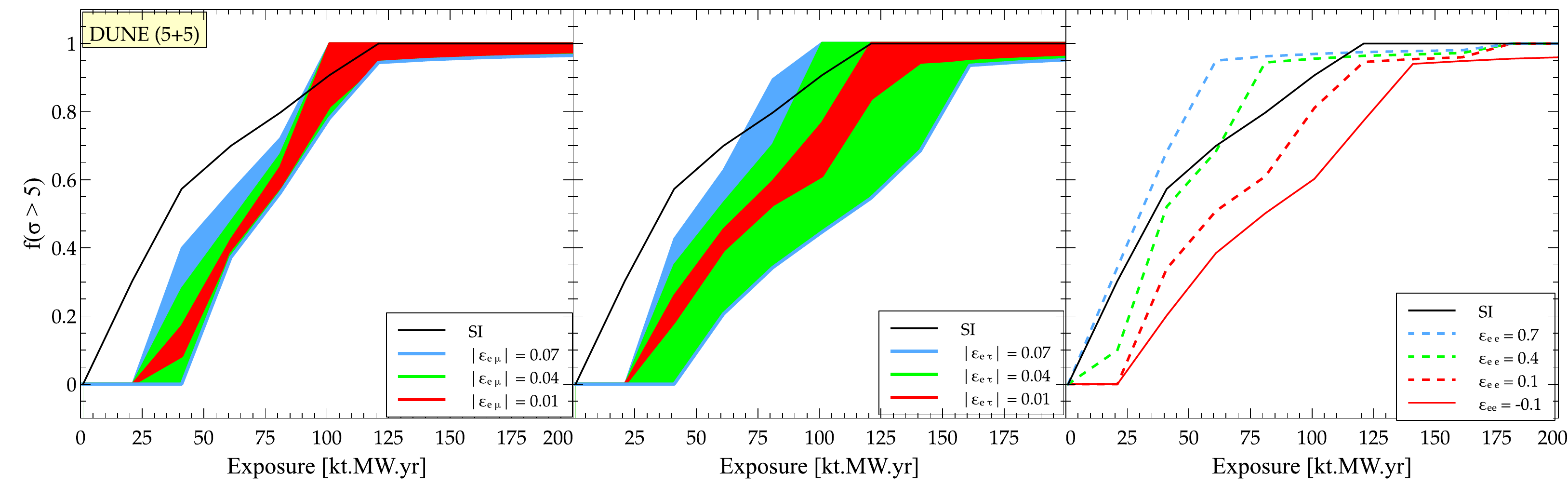}
\caption{\footnotesize{Mass ordering fraction $f^{MO} (\sigma > 5)$  plotted as a function of exposure  for \dune\, configuration and NO. 
 }}
\label{fig:exposure}
\end{figure}

 The choice of optimal exposure for  discovery of mass ordering is guided by 
the value when $f^{MO} (\sigma >  5)$ reaches its maximum value. 
 In Fig.~\ref{fig:exposure}, we show the mass ordering fraction for which the sensitivity to mass ordering
  exceeds $5 \sigma$ ($f^{MO} (\sigma > 5)$) as a function of exposure. 
For the SI case (shown as solid black  lines), we note that 
$f^{MO} (\sigma > 5)$ rises    initially  
 but saturates to its maximum value of unity  
as we go to exposures beyond $\sim 115$ kt.MW.yr.  
    So, in case of SI, the choice of optimal exposure is expected to be around  
 $ 115$ kt.MW.yr.

%-------------------------

Let us now discuss  the impact of NSI on the choice of the optimal exposure. 
  For the NSI case, the panels of Fig.~\ref{fig:ind} correspond to the three different NSI terms (taken in isolation) and we can use these to understand the main features in Fig.~\ref{fig:exposure}. There are three coloured regions (blue, green, red) 
 corresponding to the off-diagonal NSI terms in Fig.~\ref{fig:exposure} which correspond to the three values of moduli of 
 NSI parameters along with their respective phase variation 
 (analogous to the grey bands seen in Fig.~\ref{fig:ind}(a) and \ref{fig:ind}(b)). 
 For the diagonal NSI terms, there are three dashed
 lines (blue, green, red)  corresponding to three different positive values of diagonal 
 NSI parameter $\eee$ and  a solid (red) line to depict the case of $\eee < 0$  (see Fig.~\ref{fig:ind}(c)).

%----------------------

%---------------------

The  left panel in Fig.~\ref{fig:exposure} depicts the impact of $\eem$. The coloured bands stay below the solid black line mostly except around exposure of $100$ kt.MW.yr for the considered values of $\eem$. 
This is  due to  the dominating 
 statistical effect (a) and the effect (b) that operates in a small regime mentioned in Sec.~\ref{results_b}. 
For different values of $\eem$ ($|\eem|=0.01,0.04,0.07$), 
 $f^{MO} (\sigma > 5)$ gets distributed over a larger range of values   as can be seen from the  bands.   

 Incorporating the phase variation of the NSI parameter changes the value of exposure when the  
  value of $f^{MO} (\sigma > 5)$ reaches its maximum value of $\sim 1$ i.e. it changes the optimal exposure.  %
   The optimal exposure for $\eem$ varies between $\sim 100-125$ kt.MW.yr. 
  Similar effects are seen for the other off-diagonal parameter, $\eet$ which is shown in the middle panel. 
  The optimal exposure incase of $\eet$ varies between $\sim 100-150$ kt.MW.yr.

For the diagonal NSI parameter $\eee$,  $f^{MO} (\sigma > 5)$  (blue, green and red dashed lines) can be on the either side of solid black curve for a given choice of systematics (see also Fig.~\ref{fig:ind}). 
This is again due to interplay of the two kinds of effects. For negative value of this parameter, it is below the SI case and the optimal exposure shifts to $\sim 125$ kt.MW.yr. If $|\eee| = 0.7 (0.4)$, the choice of optimal exposure changes drastically from $115$ to $50 (75)$ kt.MW.yr. 
   
If we take the true ordering as IO instead of NO, the impact of NSI shown in Fig.~\ref{fig:exposure} is similar.

\subsection{Systematics}
\label{results_f}

\begin{figure}[ht!]
\centering
\includegraphics[width=\textwidth]
{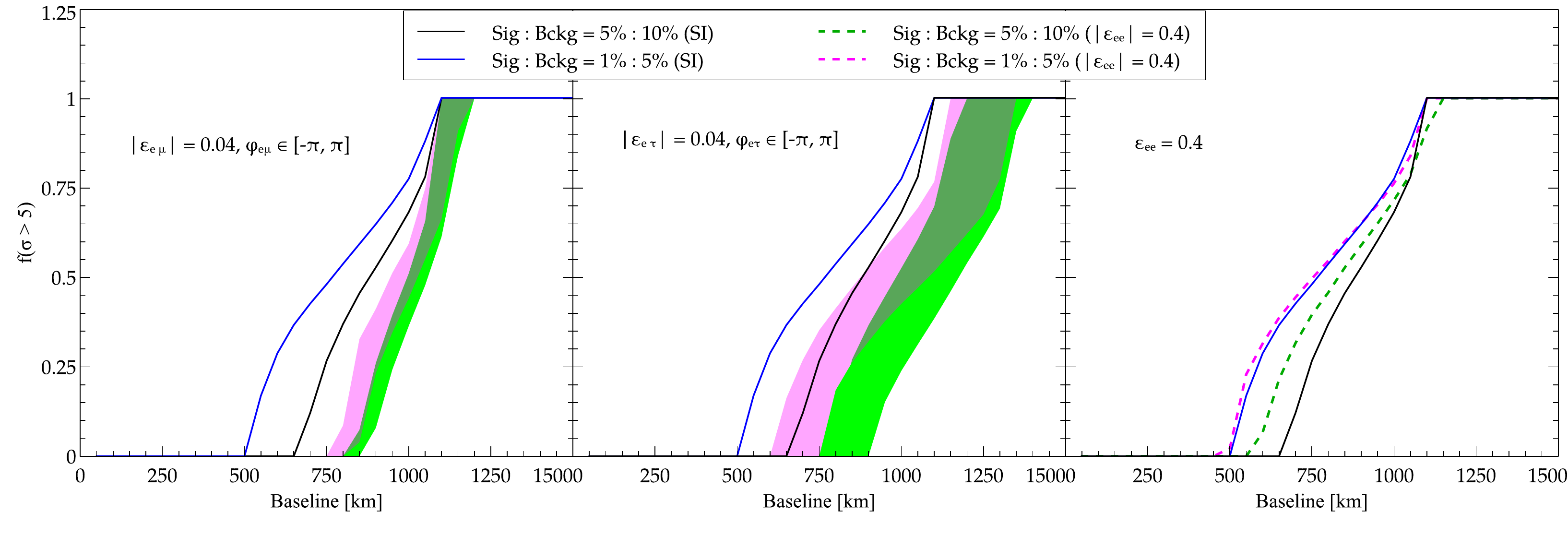}
\caption{\footnotesize{Mass ordering fraction $f^{MO} (\sigma > 5)$  plotted as a function of baseline for \dune\,  like detector configuration and NO. 
 }}
\label{fig:baseline}
\end{figure}
   The impact of different assumptions on systematics can be seen in 
  Fig.~\ref{fig:baseline}. 
The black solid curve represents our nominal choice of systematics while the blue solid curve is for an optimal choice mentioned in the legend~\cite{Bass:2013vcg}. The green (magenta) band 
corresponds to NSI case for off-diagonal parameters 
$\eem,\eet$ with full phase variation  for nominal (optimal) choice of systematics. The green (magenta) dashed curve is for $\eee$ for nominal (optimal) choice of systematics.

 %-------------------------------------------

%-------------------------------------------

It can be seen that $f^{MO} (\sigma > 5)$ nearly reaches its maximum  ($\sim 1$) possible value at around $1150$ km for SI (see Fig.~\ref{fig:baseline}). This implies that for the given 
configuration of the far detector planned for \dune, 
the optimal distance to be able to infer the mass ordering for the largest fraction of the 
 values of the CP phase is $\ge 1150$ km. 
Clearly,   in case of SI, better systematics does not significantly change the optimal baseline 
for mass ordering determination above $5\sigma$.
  For the SI case, therefore the  optimal baseline choice for mass ordering 
  sensitivity remains the same for either choice of systematics. 
 In case of NSI, the green (magenta) band shows the effect of two choices of systematics and there is an overlap between them as well as with the SI values. 
{\sl{These aspects  play a crucial role in altering the choice of best baseline for mass ordering sensitivity.}} 
 However, in presence of NSI,  for the choice of NSI phases representing the  top (bottom) edge of the green or magenta band (we have used the dashed green or magenta lines to depict the diagonal NSI terms), the optimal choice of baseline  that maximizes the mass ordering fraction changes as a function of systematics.

%-------------------------------------------
\section{Summary}
\label{sec:conclude}
%-------------------------------------------
 
 We have entered an era of precision neutrino oscillation physics. Among the  few open questions that need to be addressed, the resolution of neutrino mass ordering is an important one. Also, the unprecedented precision offered by some of the experiments allow us to probe the subdominant effects due to new  physics such as NC NSI as considered in the present work.  
 We have earlier studied the impact of NSI on the CP measurements at \dune\,  analytically as well as numerically at the level of probability and event rates in Ref.~\cite{Masud:2015xva} and at the level of  $\chi^2$ using both appearance and disappearance channels in Ref.~\cite{Masud:2016bvp}. Here we analyse the impact of NSI on the mass ordering sensitivity at the level of probability, event rates and $\chi^2$ at long baseline experiments which include \ttok, \nova\, and \dune.

Before we go on to summarise our main results, we would  like to point out that we have explicitly shown (in subsection.~\ref{framework_c} and Fig.~\ref{fig3}) that  shape of the $\chi^2$ curves for mass ordering sensitivity depends on the baseline, using statistical definition of $\chi^2$ and probability expressions for the case of SI. For NSI case, the probability expressions are  cumbersome so a simple-minded shape characterisation is not readily feasible.

 Our main conclusions are 
 \begin{itemize}
\item Among the two channels considered, $P_{\mu e}$ corresponding to the platinum channel contributes most to the sensitivity to neutrino mass ordering while $P_{\mu\mu}$ has tiny effect. The gross behaviour of sensitivity plots is dictated by $P_{\mu e}$. 
On introducing propagation NSI, 
the largest impact in the $P_{\mu e}$ channel comes from the NSI parameters $\eem$, $\eet$ and  $\eee$ and thus these parameters modify the sensitivity of long baseline experiments to neutrino mass ordering. 

 \item From the probability expressions (Eqs.~\ref{pno1}-\ref{pio4}), we notice that there is a degeneracy pertaining to
  sign of $\delta m^2_{31}$ and the CP phase $\delta$ in the case of SI. Near the peak of $P_{\mu e}$ (around 2.5 GeV) for \dune, we note that $\lambda L/2 \sim \pi/2$ and we can also note from Fig.~\ref{fig0} that the NO and IO bands are  non-overlapping around this energy. This allows us to determine the mass ordering unambiguously near the peak above $5\sigma$ irrespective  of the value of  $\delta$. This does not hold for \ttok\, or \nova\, and the ordering can be determined in favourable half plane in $\delta$ only.

\item  The dependence of mass ordering sensitivity at \dune\, on the true values of $\theta_{23}$ and 
$\delta m^2_{31}$   is shown in Fig.~\ref{fig:chisq_th23_dm31}. Both $\theta_{23}$ and $\delta m^2_{31}$ change the 
  sensitivity  for SI   as well as NSI. 
  The effect is opposite in 
  direction to CP sensitivity~\cite{Masud:2016bvp}.

\item NSI impacts the mass ordering sensitivity in two ways. The first one is the distortion in shape of the $\chi^2$ curves for SI (see Fig.~\ref{fig3}) and the second one involves the general rule laid down in Sec.~\ref{results_b} according to which there are two opposing effects at work and impact of different NSI parameters  in Fig.~\ref{fig:ind} can be understood in terms of these rules. 

\item 
 In subsection~\ref{results_d}, we  show the event rates for the appearance and disappearance channels in Figs.~\ref{fig:e1} and \ref{fig:e2} respectively  and describe the main features for the case of combined channels with and without NSI (collective NSI) for \dune.  It is shown that the presence of NSI can be inferred as modified event rates in the bins (depending on the size of NSI terms). The shape of the rate spectrum remains intact. 
We have also discussed how the  sensitivities of other long baseline experiments
 \ttok\, and \nova\, would compare with that of \dune\, for the case of collective NSI (see Figs.~\ref{fig:mo1} and \ref{fig:mo2}). 
 In case of SI, due to the smaller baselines, \ttok\, and \nova\, are not useful from the point of view of mass ordering sensitivity.   The situation worsens for \ttok\, and \nova\, when NSI effects are present. 
 The presence of  NSI worsens but does not drastically spoil the overall ability of \dune\, to resolve the neutrino
   mass ordering. The sensitivity curves in presence of NSI remain above $5 \sigma$ line for most values of 
   $\delta$ except if $\varepsilon_{ee}$ is negative.

\item Using the event rates, we depict the overlapping and non-overlapping regions for NO and IO 
considering one NSI parameter non-zero at a time in Figs.~\ref{fig:e3} and \ref{fig:e4}. The 
overlap increases as the absolute value of the NSI parameter increases for the off-diagonal 
parameters ($\varepsilon_{e\mu}$, $\varepsilon_{e\tau}$) while the overlap can either
 reduce or increase depending on the value of the diagonal parameter 
$\varepsilon_{ee}$.

\item The impact of NSI on the mass ordering fraction $f^{MO} (\sigma > 5)$ at \dune\, as a function of exposure is shown in Fig.~\ref{fig:exposure}.  The choice of optimal exposure changes in presence of NSI.

\item {{The impact of change in systematics on $f^{MO} (\sigma > 5)$ at \dune\, as a function of baseline is shown in Fig.~\ref{fig:baseline}. It is shown that the choice of optimal baseline changes when we include effects due to NSI. }}

\end{itemize}

It is shown that \dune\, is sensitive not only to the ordering of neutrino masses
  due to the SI with matter~\cite{Bass:2013vcg} but also to additional 
 NSI induced matter effects arising due to  moduli and phases of the NSI parameters. In particular, in addition to the off-diagonal NSI parameters $\eem$ and $\eet$, the diagonal NSI parameter $\eee$ also impacts the mass ordering sensitivity drastically as can be seen from Fig.~\ref{fig0}. 

In the present work, we have focussed on one of the outstanding questions in present day neutrino physics - the determination of neutrino mass ordering. 
We examine in detail the sensitivity offered by the long baseline experiments to this crucial question and how the presence of subdominant effects such as NSI spoils the sensitivity at these experiments. We have highlighted the importance of  disentangling the new physics scenario in a reliable manner to be able to make clean inference regarding mass ordering of neutrino states. This can be achieved  via a fine near detector for \dune\,~\cite{Choudhary-nd, Acciarri:2016ooe} so that the systematics be brought under control as much as possible to the extent required to cleanly determine the neutrino mass ordering and CP violation.

%%%%%%%%%%%%%%%

\section*{Acknowledgements} 
It is a pleasure to thank Raj Gandhi for useful discussions and critical comments 
on the manuscript.  
  We acknowledge the use of HRI cluster facility to carry out computations in this work.
We would like to thank the organisers of Nu Horizons VI at HRI for the 
warm hospitality during the initial stages of the present work. 
MM would like to acknowledge support from the DAE neutrino project at HRI. 
PM 
would like to thank HRI for a visit and the organisers of the Center for Theoretical Underground Physics and Related Areas (CETUP 2016) for the kind hospitality  during the finishing stages of this work. PM acknowledges support from University Grants Commission under the second phase of University with Potential of Excellence at JNU and partial support  from the European UnionÕs Horizon 2020 research and innovation programme under Marie Sklodowska-Curie grant No 674896.

\bibliographystyle{apsrev}
\bibliography{referencesnsi}

\begin{thebibliography}{10}
\expandafter\ifx\csname bibnamefont\endcsname\relax
  \def\bibnamefont#1{#1}\fi
\expandafter\ifx\csname bibfnamefont\endcsname\relax
  \def\bibfnamefont#1{#1}\fi
\expandafter\ifx\csname url\endcsname\relax
  \def\url#1{\texttt{#1}}\fi
\expandafter\ifx\csname urlprefix\endcsname\relax\def\urlprefix{URL }\fi
\providecommand{\bibinfo}[2]{#2}
\providecommand{\eprint}[2][]{\url{#2}}

\bibitem{Forero:2014bxa}
\bibinfo{author}{\bibfnamefont{D.~V.} \bibnamefont{Forero}},
  \bibinfo{author}{\bibfnamefont{M.}~\bibnamefont{Tortola}}, \bibnamefont{and}
  \bibinfo{author}{\bibfnamefont{J.~W.~F.} \bibnamefont{Valle}},
  \bibinfo{journal}{Phys. Rev.}
  \textbf{\bibinfo{volume}{D90}}(\bibinfo{number}{9}), \bibinfo{pages}{093006}
  (\bibinfo{year}{2014}), \eprint{1405.7540}.

\bibitem{Gonzalez-Garcia:2014bfa}
\bibinfo{author}{\bibfnamefont{M.~C.} \bibnamefont{Gonzalez-Garcia}},
  \bibinfo{author}{\bibfnamefont{M.}~\bibnamefont{Maltoni}}, \bibnamefont{and}
  \bibinfo{author}{\bibfnamefont{T.}~\bibnamefont{Schwetz}},
  \bibinfo{journal}{JHEP} \textbf{\bibinfo{volume}{11}}, \bibinfo{pages}{052}
  (\bibinfo{year}{2014}), \eprint{1409.5439}.

\bibitem{Albright:2006cw}
\bibinfo{author}{\bibfnamefont{C.~H.} \bibnamefont{Albright}} \bibnamefont{and}
  \bibinfo{author}{\bibfnamefont{M.-C.} \bibnamefont{Chen}},
  \bibinfo{journal}{Phys. Rev.} \textbf{\bibinfo{volume}{D74}},
  \bibinfo{pages}{113006} (\bibinfo{year}{2006}), \eprint{hep-ph/0608137}.

\bibitem{Pascoli:2005zb}
\bibinfo{author}{\bibfnamefont{S.}~\bibnamefont{Pascoli}},
  \bibinfo{author}{\bibfnamefont{S.~T.} \bibnamefont{Petcov}},
  \bibnamefont{and} \bibinfo{author}{\bibfnamefont{T.}~\bibnamefont{Schwetz}},
  \bibinfo{journal}{Nucl. Phys.} \textbf{\bibinfo{volume}{B734}},
  \bibinfo{pages}{24} (\bibinfo{year}{2006}), \eprint{hep-ph/0505226}.

\bibitem{Fukugita:1986hr}
\bibinfo{author}{\bibfnamefont{M.}~\bibnamefont{Fukugita}} \bibnamefont{and}
  \bibinfo{author}{\bibfnamefont{T.}~\bibnamefont{Yanagida}},
  \bibinfo{journal}{Phys. Lett.} \textbf{\bibinfo{volume}{B174}},
  \bibinfo{pages}{45} (\bibinfo{year}{1986}).

\bibitem{Wolfenstein:1977ue}
\bibinfo{author}{\bibfnamefont{L.}~\bibnamefont{Wolfenstein}},
  \bibinfo{journal}{Phys. Rev.} \textbf{\bibinfo{volume}{D17}},
  \bibinfo{pages}{2369} (\bibinfo{year}{1978}).

\bibitem{Mikheev:1987qk}
\bibinfo{author}{\bibfnamefont{S.~P.} \bibnamefont{Mikheev}} \bibnamefont{and}
  \bibinfo{author}{\bibfnamefont{A.~Y.} \bibnamefont{Smirnov}},
  \bibinfo{journal}{Sov. Phys. Usp.} \textbf{\bibinfo{volume}{30}},
  \bibinfo{pages}{759} (\bibinfo{year}{1987}).

\bibitem{Ge:2016xya}
\bibinfo{author}{\bibfnamefont{S.-F.} \bibnamefont{Ge}},
  \bibinfo{author}{\bibfnamefont{P.}~\bibnamefont{Pasquini}},
  \bibinfo{author}{\bibfnamefont{M.}~\bibnamefont{Tortola}}, \bibnamefont{and}
  \bibinfo{author}{\bibfnamefont{J.~W.~F.} \bibnamefont{Valle}}
  (\bibinfo{year}{2016}), \eprint{1605.01670}.

\bibitem{Ohlsson:2012kf}
\bibinfo{author}{\bibfnamefont{T.}~\bibnamefont{Ohlsson}},
  \bibinfo{journal}{Rept. Prog. Phys.} \textbf{\bibinfo{volume}{76}},
  \bibinfo{pages}{044201} (\bibinfo{year}{2013}), \eprint{1209.2710}.

\bibitem{Miranda:2015dra}
\bibinfo{author}{\bibfnamefont{O.~G.} \bibnamefont{Miranda}} \bibnamefont{and}
  \bibinfo{author}{\bibfnamefont{H.}~\bibnamefont{Nunokawa}},
  \bibinfo{journal}{New J. Phys.}
  \textbf{\bibinfo{volume}{17}}(\bibinfo{number}{9}), \bibinfo{pages}{095002}
  (\bibinfo{year}{2015}), \eprint{1505.06254}.

\bibitem{Marciano:2006uc}
\bibinfo{author}{\bibfnamefont{W.}~\bibnamefont{Marciano}} \bibnamefont{and}
  \bibinfo{author}{\bibfnamefont{Z.}~\bibnamefont{Parsa}},
  \bibinfo{journal}{Nucl. Phys. Proc. Suppl.} \textbf{\bibinfo{volume}{221}},
  \bibinfo{pages}{166} (\bibinfo{year}{2011}), \eprint{hep-ph/0610258}.

\bibitem{Bass:2013vcg}
\bibinfo{author}{\bibfnamefont{M.}~\bibnamefont{Bass}} \emph{et~al.}
  (\bibinfo{collaboration}{LBNE Collaboration}), \bibinfo{journal}{Phys. Rev.}
  \textbf{\bibinfo{volume}{D91}}, \bibinfo{pages}{052015}
  (\bibinfo{year}{2015}), \eprint{1311.0212}.

\bibitem{Acciarri:2015uup}
\bibinfo{author}{\bibfnamefont{R.}~\bibnamefont{Acciarri}} \emph{et~al.}
  (\bibinfo{collaboration}{DUNE})  (\bibinfo{year}{2015}), \eprint{1512.06148}.

\bibitem{Acciarri:2016ooe}
\bibinfo{author}{\bibfnamefont{R.}~\bibnamefont{Acciarri}} \emph{et~al.}
  (\bibinfo{collaboration}{DUNE})  (\bibinfo{year}{2016}), \eprint{1601.02984}.

\bibitem{Acciarri:2016crz}
\bibinfo{author}{\bibfnamefont{R.}~\bibnamefont{Acciarri}} \emph{et~al.}
  (\bibinfo{collaboration}{DUNE})  (\bibinfo{year}{2016}), \eprint{1601.05471}.

\bibitem{Masud:2015xva}
\bibinfo{author}{\bibfnamefont{M.}~\bibnamefont{Masud}},
  \bibinfo{author}{\bibfnamefont{A.}~\bibnamefont{Chatterjee}},
  \bibnamefont{and} \bibinfo{author}{\bibfnamefont{P.}~\bibnamefont{Mehta}}
  (\bibinfo{year}{2015}), \eprint{1510.08261}.

\bibitem{Coloma:2015kiu}
\bibinfo{author}{\bibfnamefont{P.}~\bibnamefont{Coloma}},
  \bibinfo{journal}{JHEP} \textbf{\bibinfo{volume}{03}}, \bibinfo{pages}{016}
  (\bibinfo{year}{2016}), \eprint{1511.06357}.

\bibitem{deGouvea:2015ndi}
\bibinfo{author}{\bibfnamefont{A.}~\bibnamefont{de~Gouva}} \bibnamefont{and}
  \bibinfo{author}{\bibfnamefont{K.~J.} \bibnamefont{Kelly}},
  \bibinfo{journal}{Nucl. Phys.} \textbf{\bibinfo{volume}{B908}},
  \bibinfo{pages}{318} (\bibinfo{year}{2016}), \eprint{1511.05562}.

\bibitem{Forero:2016cmb}
\bibinfo{author}{\bibfnamefont{D.~V.} \bibnamefont{Forero}} \bibnamefont{and}
  \bibinfo{author}{\bibfnamefont{P.}~\bibnamefont{Huber}}
  (\bibinfo{year}{2016}), \eprint{1601.03736}.

\bibitem{Liao:2016hsa}
\bibinfo{author}{\bibfnamefont{J.}~\bibnamefont{Liao}},
  \bibinfo{author}{\bibfnamefont{D.}~\bibnamefont{Marfatia}}, \bibnamefont{and}
  \bibinfo{author}{\bibfnamefont{K.}~\bibnamefont{Whisnant}},
  \bibinfo{journal}{Phys. Rev.}
  \textbf{\bibinfo{volume}{D93}}(\bibinfo{number}{9}), \bibinfo{pages}{093016}
  (\bibinfo{year}{2016}), \eprint{1601.00927}.

\bibitem{Huitu:2016bmb}
\bibinfo{author}{\bibfnamefont{K.}~\bibnamefont{Huitu}},
  \bibinfo{author}{\bibfnamefont{T.~J.} \bibnamefont{KŠrkkŠinen}},
  \bibinfo{author}{\bibfnamefont{J.}~\bibnamefont{Maalampi}}, \bibnamefont{and}
  \bibinfo{author}{\bibfnamefont{S.}~\bibnamefont{Vihonen}},
  \bibinfo{journal}{Phys. Rev.}
  \textbf{\bibinfo{volume}{D93}}(\bibinfo{number}{5}), \bibinfo{pages}{053016}
  (\bibinfo{year}{2016}), \eprint{1601.07730}.

\bibitem{Bakhti:2016prn}
\bibinfo{author}{\bibfnamefont{P.}~\bibnamefont{Bakhti}} \bibnamefont{and}
  \bibinfo{author}{\bibfnamefont{Y.}~\bibnamefont{Farzan}}
  (\bibinfo{year}{2016}), \eprint{1602.07099}.

\bibitem{Coloma:2016gei}
\bibinfo{author}{\bibfnamefont{P.}~\bibnamefont{Coloma}} \bibnamefont{and}
  \bibinfo{author}{\bibfnamefont{T.}~\bibnamefont{Schwetz}}
  (\bibinfo{year}{2016}), \eprint{1604.05772}.

\bibitem{Masud:2016bvp}
\bibinfo{author}{\bibfnamefont{M.}~\bibnamefont{Masud}} \bibnamefont{and}
  \bibinfo{author}{\bibfnamefont{P.}~\bibnamefont{Mehta}}
  (\bibinfo{year}{2016}), \eprint{1603.01380}.

\bibitem{TheT2KCollaboration01042015}
\bibinfo{author}{\bibfnamefont{K.}~\bibnamefont{Abe}} \emph{et~al.}
  (\bibinfo{collaboration}{T2K}), \bibinfo{journal}{Progress of Theoretical and
  Experimental Physics} \textbf{\bibinfo{volume}{2015}}(\bibinfo{number}{4})
  (\bibinfo{year}{2015}).

\bibitem{nova}
\bibinfo{author}{\bibfnamefont{D.~S.} \bibnamefont{Ayres}} \emph{et~al.}
  (\bibinfo{collaboration}{NOvA Collaboration}), \bibinfo{journal}{Arxiv
  eprints}  (\bibinfo{year}{2004}), \eprint{hep-ex/0503053}.

\bibitem{Adamson:2016tbq}
\bibinfo{author}{\bibfnamefont{P.}~\bibnamefont{Adamson}} \emph{et~al.}
  (\bibinfo{collaboration}{NOvA})  (\bibinfo{year}{2016}), \eprint{1601.05022}.

\bibitem{Abe:2015zbg}
\bibinfo{author}{\bibfnamefont{K.}~\bibnamefont{Abe}} \emph{et~al.}
  (\bibinfo{collaboration}{Hyper-Kamiokande Proto-Collaboration}),
  \bibinfo{journal}{PTEP} \textbf{\bibinfo{volume}{2015}},
  \bibinfo{pages}{053C02} (\bibinfo{year}{2015}), \eprint{1502.05199}.

\bibitem{Djurcic:2015vqa}
\bibinfo{author}{\bibfnamefont{Z.}~\bibnamefont{Djurcic}} \emph{et~al.}
  (\bibinfo{collaboration}{JUNO})  (\bibinfo{year}{2015}), \eprint{1508.07166}.

\bibitem{An:2015jdp}
\bibinfo{author}{\bibfnamefont{F.}~\bibnamefont{An}} \emph{et~al.}
  (\bibinfo{collaboration}{JUNO}), \bibinfo{journal}{J. Phys.}
  \textbf{\bibinfo{volume}{G43}}(\bibinfo{number}{3}), \bibinfo{pages}{030401}
  (\bibinfo{year}{2016}), \eprint{1507.05613}.

\bibitem{Ahmed:2015jtv}
\bibinfo{author}{\bibfnamefont{S.}~\bibnamefont{Ahmed}} \emph{et~al.}
  (\bibinfo{collaboration}{ICAL})  (\bibinfo{year}{2015}), \eprint{1505.07380}.

\bibitem{Adrian-Martinez:2016fdl}
\bibinfo{author}{\bibfnamefont{S.}~\bibnamefont{Adrian-Martinez}} \emph{et~al.}
  (\bibinfo{collaboration}{KM3Net})  (\bibinfo{year}{2016}),
  \eprint{1601.07459}.

\bibitem{Aartsen:2014oha}
\bibinfo{author}{\bibfnamefont{M.~G.} \bibnamefont{Aartsen}} \emph{et~al.}
  (\bibinfo{collaboration}{IceCube PINGU})  (\bibinfo{year}{2014}),
  \eprint{1401.2046}.

\bibitem{Ribordy:2013set}
\bibinfo{author}{\bibfnamefont{M.}~\bibnamefont{Ribordy}} \bibnamefont{and}
  \bibinfo{author}{\bibfnamefont{A.~Y.} \bibnamefont{Smirnov}},
  \bibinfo{journal}{Phys. Rev.}
  \textbf{\bibinfo{volume}{D87}}(\bibinfo{number}{11}), \bibinfo{pages}{113007}
  (\bibinfo{year}{2013}), \eprint{1303.0758}.

\bibitem{Akhmedov:2012ah}
\bibinfo{author}{\bibfnamefont{E.~K.} \bibnamefont{Akhmedov}},
  \bibinfo{author}{\bibfnamefont{S.}~\bibnamefont{Razzaque}}, \bibnamefont{and}
  \bibinfo{author}{\bibfnamefont{A.~Y.} \bibnamefont{Smirnov}}
  (\bibinfo{year}{2012}), \eprint{1205.7071}.

\bibitem{Hannestad:2016fog}
\bibinfo{author}{\bibfnamefont{S.}~\bibnamefont{Hannestad}} \bibnamefont{and}
  \bibinfo{author}{\bibfnamefont{T.}~\bibnamefont{Schwetz}}
  (\bibinfo{year}{2016}), \eprint{1606.04691}.

\bibitem{Antusch:2008tz}
\bibinfo{author}{\bibfnamefont{S.}~\bibnamefont{Antusch}},
  \bibinfo{author}{\bibfnamefont{J.~P.} \bibnamefont{Baumann}},
  \bibnamefont{and}
  \bibinfo{author}{\bibfnamefont{E.}~\bibnamefont{Fernandez-Martinez}},
  \bibinfo{journal}{Nucl. Phys.} \textbf{\bibinfo{volume}{B810}},
  \bibinfo{pages}{369} (\bibinfo{year}{2009}), \eprint{0807.1003}.

\bibitem{Farzan:2015doa}
\bibinfo{author}{\bibfnamefont{Y.}~\bibnamefont{Farzan}},
  \bibinfo{journal}{Phys. Lett.} \textbf{\bibinfo{volume}{B748}},
  \bibinfo{pages}{311} (\bibinfo{year}{2015}), \eprint{1505.06906}.

\bibitem{Farzan:2015hkd}
\bibinfo{author}{\bibfnamefont{Y.}~\bibnamefont{Farzan}} \bibnamefont{and}
  \bibinfo{author}{\bibfnamefont{I.~M.} \bibnamefont{Shoemaker}}
  (\bibinfo{year}{2015}), \eprint{1512.09147}.

\bibitem{Qian:2015waa}
\bibinfo{author}{\bibfnamefont{X.}~\bibnamefont{Qian}} \bibnamefont{and}
  \bibinfo{author}{\bibfnamefont{P.}~\bibnamefont{Vogel}},
  \bibinfo{journal}{Prog. Part. Nucl. Phys.} \textbf{\bibinfo{volume}{83}},
  \bibinfo{pages}{1} (\bibinfo{year}{2015}), \eprint{1505.01891}.

\bibitem{Rashed:2016rda}
\bibinfo{author}{\bibfnamefont{A.}~\bibnamefont{Rashed}} \bibnamefont{and}
  \bibinfo{author}{\bibfnamefont{A.}~\bibnamefont{Datta}}
  (\bibinfo{year}{2016}), \eprint{1603.09031}.

\bibitem{Beringer:1900zz}
\bibinfo{author}{\bibfnamefont{J.}~\bibnamefont{Beringer}} \emph{et~al.}
  (\bibinfo{collaboration}{Particle Data Group}), \bibinfo{journal}{Phys. Rev.}
  \textbf{\bibinfo{volume}{D86}}, \bibinfo{pages}{010001}
  (\bibinfo{year}{2012}).

\bibitem{Chatterjee:2014gxa}
\bibinfo{author}{\bibfnamefont{A.}~\bibnamefont{Chatterjee}},
  \bibinfo{author}{\bibfnamefont{P.}~\bibnamefont{Mehta}},
  \bibinfo{author}{\bibfnamefont{D.}~\bibnamefont{Choudhury}},
  \bibnamefont{and} \bibinfo{author}{\bibfnamefont{R.}~\bibnamefont{Gandhi}},
  \bibinfo{journal}{Phys. Rev.}
  \textbf{\bibinfo{volume}{D93}}(\bibinfo{number}{9}), \bibinfo{pages}{093017}
  (\bibinfo{year}{2016}), \eprint{1409.8472}.

\bibitem{Choubey:2015xha}
\bibinfo{author}{\bibfnamefont{S.}~\bibnamefont{Choubey}},
  \bibinfo{author}{\bibfnamefont{A.}~\bibnamefont{Ghosh}},
  \bibinfo{author}{\bibfnamefont{T.}~\bibnamefont{Ohlsson}}, \bibnamefont{and}
  \bibinfo{author}{\bibfnamefont{D.}~\bibnamefont{Tiwari}},
  \bibinfo{journal}{JHEP} \textbf{\bibinfo{volume}{12}}, \bibinfo{pages}{126}
  (\bibinfo{year}{2015}), \eprint{1507.02211}.

\bibitem{Akhmedov:2004ny}
\bibinfo{author}{\bibfnamefont{E.~K.} \bibnamefont{Akhmedov}},
  \bibinfo{author}{\bibfnamefont{R.}~\bibnamefont{Johansson}},
  \bibinfo{author}{\bibfnamefont{M.}~\bibnamefont{Lindner}},
  \bibinfo{author}{\bibfnamefont{T.}~\bibnamefont{Ohlsson}}, \bibnamefont{and}
  \bibinfo{author}{\bibfnamefont{T.}~\bibnamefont{Schwetz}},
  \bibinfo{journal}{JHEP} \textbf{\bibinfo{volume}{0404}}, \bibinfo{pages}{078}
  (\bibinfo{year}{2004}), \eprint{hep-ph/0402175}.

\bibitem{Prakash:2012az}
\bibinfo{author}{\bibfnamefont{S.}~\bibnamefont{Prakash}},
  \bibinfo{author}{\bibfnamefont{S.~K.} \bibnamefont{Raut}}, \bibnamefont{and}
  \bibinfo{author}{\bibfnamefont{S.~U.} \bibnamefont{Sankar}},
  \bibinfo{journal}{Phys. Rev.} \textbf{\bibinfo{volume}{D86}},
  \bibinfo{pages}{033012} (\bibinfo{year}{2012}), \eprint{1201.6485}.

\bibitem{Huber:2004ka}
\bibinfo{author}{\bibfnamefont{P.}~\bibnamefont{Huber}},
  \bibinfo{author}{\bibfnamefont{M.}~\bibnamefont{Lindner}}, \bibnamefont{and}
  \bibinfo{author}{\bibfnamefont{W.}~\bibnamefont{Winter}},
  \bibinfo{journal}{Comput. Phys. Commun.} \textbf{\bibinfo{volume}{167}},
  \bibinfo{pages}{195} (\bibinfo{year}{2005}), \eprint{hep-ph/0407333}.

\bibitem{Kopp:2006wp}
\bibinfo{author}{\bibfnamefont{J.}~\bibnamefont{Kopp}}, \bibinfo{journal}{Int.
  J. Mod. Phys.} \textbf{\bibinfo{volume}{C19}}, \bibinfo{pages}{523}
  (\bibinfo{year}{2008}), \eprint{physics/0610206}.

\bibitem{Huber:2007ji}
\bibinfo{author}{\bibfnamefont{P.}~\bibnamefont{Huber}},
  \bibinfo{author}{\bibfnamefont{J.}~\bibnamefont{Kopp}},
  \bibinfo{author}{\bibfnamefont{M.}~\bibnamefont{Lindner}},
  \bibinfo{author}{\bibfnamefont{M.}~\bibnamefont{Rolinec}}, \bibnamefont{and}
  \bibinfo{author}{\bibfnamefont{W.}~\bibnamefont{Winter}},
  \bibinfo{journal}{Comput. Phys. Commun.} \textbf{\bibinfo{volume}{177}},
  \bibinfo{pages}{432} (\bibinfo{year}{2007}), \eprint{hep-ph/0701187}.

\bibitem{Kopp:2007ne}
\bibinfo{author}{\bibfnamefont{J.}~\bibnamefont{Kopp}},
  \bibinfo{author}{\bibfnamefont{M.}~\bibnamefont{Lindner}},
  \bibinfo{author}{\bibfnamefont{T.}~\bibnamefont{Ota}}, \bibnamefont{and}
  \bibinfo{author}{\bibfnamefont{J.}~\bibnamefont{Sato}},
  \bibinfo{journal}{Phys. Rev.} \textbf{\bibinfo{volume}{D77}},
  \bibinfo{pages}{013007} (\bibinfo{year}{2008}), \eprint{0708.0152}.

\bibitem{Dziewonski:1981xy}
\bibinfo{author}{\bibfnamefont{A.~M.} \bibnamefont{Dziewonski}}
  \bibnamefont{and} \bibinfo{author}{\bibfnamefont{D.~L.}
  \bibnamefont{Anderson}}, \bibinfo{journal}{Phys. Earth Planet. Interiors}
  \textbf{\bibinfo{volume}{25}}, \bibinfo{pages}{297} (\bibinfo{year}{1981}).

\bibitem{Gandhi:2004bj}
\bibinfo{author}{\bibfnamefont{R.}~\bibnamefont{Gandhi}},
  \bibinfo{author}{\bibfnamefont{P.}~\bibnamefont{Ghoshal}},
  \bibinfo{author}{\bibfnamefont{S.}~\bibnamefont{Goswami}},
  \bibinfo{author}{\bibfnamefont{P.}~\bibnamefont{Mehta}}, \bibnamefont{and}
  \bibinfo{author}{\bibfnamefont{S.~U.} \bibnamefont{Sankar}},
  \bibinfo{journal}{Phys.Rev.} \textbf{\bibinfo{volume}{D73}},
  \bibinfo{pages}{053001} (\bibinfo{year}{2006}), \eprint{hep-ph/0411252}.

\bibitem{GonzalezGarcia:2012sz}
\bibinfo{author}{\bibfnamefont{M.}~\bibnamefont{Gonzalez-Garcia}},
  \bibinfo{author}{\bibfnamefont{M.}~\bibnamefont{Maltoni}},
  \bibinfo{author}{\bibfnamefont{J.}~\bibnamefont{Salvado}}, \bibnamefont{and}
  \bibinfo{author}{\bibfnamefont{T.}~\bibnamefont{Schwetz}},
  \bibinfo{journal}{JHEP} \textbf{\bibinfo{volume}{1212}}, \bibinfo{pages}{123}
  (\bibinfo{year}{2012}), \eprint{1209.3023}.

\bibitem{Capozzi:2013csa}
\bibinfo{author}{\bibfnamefont{F.}~\bibnamefont{Capozzi}},
  \bibinfo{author}{\bibfnamefont{G.}~\bibnamefont{Fogli}},
  \bibinfo{author}{\bibfnamefont{E.}~\bibnamefont{Lisi}},
  \bibinfo{author}{\bibfnamefont{A.}~\bibnamefont{Marrone}},
  \bibinfo{author}{\bibfnamefont{D.}~\bibnamefont{Montanino}}, \emph{et~al.},
  \bibinfo{journal}{Phys.Rev.} \textbf{\bibinfo{volume}{D89}},
  \bibinfo{pages}{093018} (\bibinfo{year}{2014}), \eprint{1312.2878}.

\bibitem{Qian:2012zn}
\bibinfo{author}{\bibfnamefont{X.}~\bibnamefont{Qian}},
  \bibinfo{author}{\bibfnamefont{A.}~\bibnamefont{Tan}},
  \bibinfo{author}{\bibfnamefont{W.}~\bibnamefont{Wang}},
  \bibinfo{author}{\bibfnamefont{J.~J.} \bibnamefont{Ling}},
  \bibinfo{author}{\bibfnamefont{R.~D.} \bibnamefont{McKeown}},
  \bibnamefont{and} \bibinfo{author}{\bibfnamefont{C.}~\bibnamefont{Zhang}},
  \bibinfo{journal}{Phys. Rev.} \textbf{\bibinfo{volume}{D86}},
  \bibinfo{pages}{113011} (\bibinfo{year}{2012}), \eprint{1210.3651}.

\bibitem{Barger:2014dfa}
\bibinfo{author}{\bibfnamefont{V.}~\bibnamefont{Barger}},
  \bibinfo{author}{\bibfnamefont{A.}~\bibnamefont{Bhattacharya}},
  \bibinfo{author}{\bibfnamefont{A.}~\bibnamefont{Chatterjee}},
  \bibinfo{author}{\bibfnamefont{R.}~\bibnamefont{Gandhi}},
  \bibinfo{author}{\bibfnamefont{D.}~\bibnamefont{Marfatia}}, \bibnamefont{and}
  \bibinfo{author}{\bibfnamefont{M.}~\bibnamefont{Masud}},
  \bibinfo{journal}{Int. J. Mod. Phys.}
  \textbf{\bibinfo{volume}{A31}}(\bibinfo{number}{07}),
  \bibinfo{pages}{1650020} (\bibinfo{year}{2016}), \eprint{1405.1054}.

\bibitem{Messier:1999kj}
\bibinfo{author}{\bibfnamefont{M.~D.} \bibnamefont{Messier}},
  \emph{\bibinfo{title}{{Evidence for neutrino mass from observations of
  atmospheric neutrinos with Super-Kamiokande}}}, Ph.D. thesis,
  \bibinfo{school}{Boston U.} (\bibinfo{year}{1999}),
  \urlprefix\url{http://wwwlib.umi.com/dissertations/fullcit?p9923965}.

\bibitem{Choudhary-nd}
\bibinfo{author}{\bibnamefont{{B. Choudhary, R. Gandhi, S. R. Mishra, Shekhar
  Mishra and J. Strait}}}, \emph{\bibinfo{title}{{Proposal of Indian
  Institutions and Fermilab Collaboration for Participation in the
  Long-Baseline Neutrino Experiment at Fermilab}}},
  \bibinfo{howpublished}{\url{http://lbne2-docdb.fnal.gov/cgi-bin/RetrieveFile?docid=6704&filename=LBNE-India-DPR-V12-Science.pdf&version=1}}
  (\bibinfo{year}{2012}).

\end{thebibliography}

\end{document}